%% file: paper.tex
\documentclass[letterpaper,twocolumn,10pt]{article}
\usepackage{usenix2019_v3}
\usepackage{formatting/shortcuts}

\begin{document}

\date{}
\pagestyle{empty}

\title{\Large \bf BesFS: A POSIX Filesystem for Enclaves with a Mechanized Safety Proof}

\author{
{\rm Shweta Shinde}
\thanks{These joint first authors contributed equally to this work.}
~\thanks{Part of the research was done while at National University of Singapore.}\\
{\normalsize University of California, Berkeley}
\and
{\rm Shengyi Wang\footnotemark[1]}\\
{\normalsize National University of Singapore}
\and
{\rm Pinghai Yuan}\\
{\normalsize National University of Singapore}
\and
{\rm Aquinas Hobor}\\
{\normalsize National University of Singapore}\\ 
{\normalsize \& Yale-NUS College} \\
\and
{\rm Abhik Roychoudhury}\\
{\normalsize National University of Singapore}
\and
{\rm Prateek Saxena}\\
{\normalsize National University of Singapore}
}  

\maketitle

\begin{abstract}
\input{chapters/abstract}
\end{abstract}

\input{paper-body}

{\small%
\bibliographystyle{abbrv}
\def\UrlBreaks{\do\/\do-}
\bibliography{paper}
}
\end{document}

%% file: chapters/abstract.tex
New trusted computing primitives such as Intel SGX have shown the
feasibility of running user-level applications in enclaves on a
commodity trusted processor without trusting a large OS. However, the
OS can still compromise the integrity of an enclave by tampering with
the system call return values. In fact, it has been shown that a
subclass of these attacks, called Iago attacks, enables arbitrary
logic execution in enclave programs. Existing enclave systems have
very large TCB and they implement ad-hoc checks at the system call
interface which are hard to verify for completeness. To this end, we
present \codename---the first filesystem interface which provably
protects the enclave  integrity against a completely malicious OS. We
prove $167$ lemmas and $2$ key theorems in $4625$ lines of \coq proof
scripts, which directly proves the safety properties of the \codename
specification. \codename comprises of $15$ APIs with compositional
safety and is expressive enough to support $31$ real applications we
test. \codename integrates into existing SGX-enabled applications with
minimal impact to TCB. \codename can serve as a reference
implementation for hand-coded API checks. 

%% file: paper-body.tex
\input{chapters/intro}
\input{chapters/problem}
\input{chapters/design}
\input{chapters/impl}
\input{chapters/safety-proof}

\input{chapters/coq-to-exec}
\input{chapters/eval}
\input{chapters/related}
\input{chapters/discussion}
\input{chapters/conclusion}
\input{chapters/ack}
\input{chapters/availability}

%% file: chapters/intro.tex
\section{Introduction}
\label{sec:intro}

Existing computer systems encompass millions of lines of complex
operating system (OS) code, which is highly susceptible to
vulnerabilities, but is trusted by all user-level applications. In the
last decade, a line of research has established that trusting an OS
implementation is {\em not} necessary. Specifically, new trusted
computing primitives (e.g., Intel SGX~\cite{sgx},
Sanctum~\cite{sanctum}, Keystone~\cite{keystone}) have shown the
feasibility of running user-level applications on a commodity trusted
processor without trusting a large OS. These are called {\em enclaved
execution} primitives, using the parlance introduced by Intel SGX---a
widely shipping feature in commodity Intel processors today.
Applications on such systems run isolated from the OS in CPU-protected
memory regions called enclaves; with various adversary models
supported in individual designs~\cite{keystone, sgx, komodo-sosp17,
podarch, sanctum}.

Enclave systems promise to minimize the trusted code base
(TCB) of a security-critical application. Ideally, the TCB can be made
boiler-plate and small enough to be {\em formally verified} to be free
of vulnerabilities. Towards this vision, recent works have formally
specified and checked the interfaces between the enclave and the
CPU~\cite{tap-ccs17, komodo-sosp17}, as well as verified enclave
confidentiality properties~\cite{moat, slashconfidential}. One
critical gap remains unaddressed: verifying the integrity of the
application from a {\em hostile} OS. Applications are increasingly
becoming easier to port to enclaves~\cite{panoply,haven,graphene-sgx,
scone}; however, these legacy applications optimistically assume that
the OS is benign. A hostile OS, however, can behave arbitrarily by
violating assumptions inherent in the basic abstractions of processes
or files and exchange malicious data with the application. This
well-known attack was originally identified by Ports and Garfinkel as
{\em system call tampering}~\cite{ports-garfinkel}, more recently
discussed as a subclass called Iago attacks~\cite{iago}.

A number of enclave execution platforms have recognized this channel
of attack but left specifying the necessary checks out of scope. For
instance, systems such as Haven~\cite{haven}, Google
Asylo~\cite{asylo}, Microsoft Open Enclave~\cite{open-enclave}, Intel
SGX SDK~\cite{intel-fs}, \panoply~\cite{panoply},
Graphene-SGX~\cite{graphene-sgx}, and Scone~\cite{scone} built on
Intel SGX have alluded to syscall tampering defense as an important
challenge; however, none of these systems claim a guaranteed defense.
One of the reasons is that a hostile OS can deviate from the intended
behavior in so many ways. Reasoning about a {\em complete} set of
checks that suffice to capture all attacks is difficult.

In this work, we take a step towards a formally verified TCB to
protect the integrity of enclaves against a hostile OS. To maximize the
eliminated attack surface and compatibility with existing OSes, we
propose to safeguard at the POSIX system call interface. We scope this
work to the filesystem subset of the POSIX API. Our main contribution
is \codename---a POSIX-compliant filesystem specification with formal
guarantees of integrity and a machine-checked proof of its
implementation in a high-level language. Client applications running
in SGX enclaves interact with a commodity (e.g., Linux) OS via our
\codename implementation, running as a library (see
Figure~\ref{fig:overview}). Applications use the POSIX filesystem API 
transparently (see Table~\ref{tab:api}), requiring minimal integration
changes. Being formally verified, \codename specifications and
implementation can further be used to test or verify other
implementations based on SGX and similar primitives.

\paragraph{Challenges \& Approach}. 
The main set of challenges in developing \codename are two-fold. The
first challenge is in establishing the ``right'' specification of the
filesystem interface, such that it is both safe (captures well-known
attacks) and admits common benign functionality. To show safety, we
outline various known syscall tampering attacks and prove that
\codename interface specification defeats at least these attacks by
its very design. The attacks defeated are not limited to identified
list here---in fact, any deviations from the defined behavior of the
\codename interface is treated as a violation, aborting the client
program safely. To address compatibility, we empirically test a wide
variety of   real-world applications and benchmarks with a
\codename-enhanced system for running SGX applications. These tests
show a modest impact on compatibility, showing that the \codename
specification is rich enough to run many practical applications on
commodity OS implementations. The \codename API has only $15$ core
operations. However, it is accompanied crucially by a composition
theorem that safeguards chaining all combinations of operations,
making extensions to high-level APIs (e.g., \libc) easy.

The second challenge is in the execution of the proof of the \codename
implementation itself. Our proof turns out to be challenging because
the properties require higher-order logic (hence the need for \coq)
and reasoning about {\em arbitrary} behavior at points at which the OS
is invoked. Specifically, the filesystem is modeled as a
state-transition system where each filesystem operation transitions
from one state to another. Various design challenges arise
(Section~\ref{sec:proof}) in handling a stateful implementation in the
stateless proof system of \coq and uncovering inductive proof
strategies for recursive data structures used in the \codename
implementation. These proof strategies are more involved than \coq's
automatic tactics.

\codename is specified, implemented and formally verified in \coq
which is a higher-level language. Converting \coq code to machine code
is out of the scope of this paper. Most existing systems do not
provide these guarantees even for non-enclave code. There are several
intermediate challenges in such a conversion, especially when it is
enclave-bound. Thus we resort to a hand-coded conversion of \codename
implementation from \coq-to-$\tt{C}$ and then use an Intel SGX
compatible compiler to obtain machine code which can execute inside
the enclave. For the completeness of the paper, we outline various
challenges we faced in our attempt to generate enclave-bound machine
code from our \coq implementation of \codename. We discuss the
existing alternatives and the required additions to the immediate
state of systems to make this feasible. %

\paragraph{Results}. 
Our \codename \coq proof comprises of $167$ theorems and $4625$ LOC.
Our hand-coded $\tt{C}$ implementation of \codename is $1449$ LOC and
we add $724$ LOC of stubs for compatibility with enclave code. We use
this $\tt{C}$ implementation for our performance evaluation. We
demonstrate the expressiveness of \codename by supporting $31$
programs from benchmarks and real-world applications. We show that
\codename is compatible with state-of-the-art filesystems, benchmarks,
and applications we tested. It aids in finding implementation mistakes
in existing filesystem APIs exposed by Intel SGX frameworks. We hope
\codename will serve as a specification for future optimizations and
other hand-coded implementations. 

\paragraph{Contributions.}
We make the following contributions:
\begin{itemize}
\squish
	
\item 
We formally model the class of attacks that the OS can launch against
SGX enclaves via the filesystem API and develop a complete set of 
specifications to disable them. 

\item 
We present \codename---a formally verified set of API implementations
in \coq which are machine-checked for their soundness w.r.t. the API
specifications. Our auto-generated run-time monitoring mechanism
ensures that the concrete filesystem execution stays within the
envelope of our specification. 

\item 
We prove $167$ lemmas and $2$ key theorems in $4625$ LOC \coq. We
evaluate correctness, compatibility, and expressiveness of \codename.
We showcase \codename on $31$ programs from real-world applications 
and standard benchmarks for CPU, I/O, and filesystem workloads.

\end{itemize}

%% file: chapters/problem.tex
\section{Problem}
\label{sec:problem}

There has been long-standing research on protecting the OS from
user-level applications. In this work, the threat model is reversed;
the applications demand protection against a malicious OS kernel. We
briefly review Intel SGX specifics and highlight the need for a formal
approach.

\subsection{Background \& Setup}
\label{sec:background}

Intel SGX provides a set of CPU instructions which can protect
selected parts of user-level application logic from an untrusted
operating system. Specifically, the developer can encapsulate
sensitive logic inside an {\em enclave}. The CPU allocates protected
physical memory from {\em Enclave Page Cache (EPC)} that backs the
enclave main memory and its content is encrypted. Only the owner
enclave can access its own content at any point during execution. The
hardware does not allow any other process or the OS to modify code and
data or read plain text inside the enclave boundary. Interested
readers can refer to ~\cite{sgx-explained} for full details.

Due to the strict memory protection, unprotected instructions such as
$\tt{syscall}$ are illegal inside the enclave. However, the
application can use {\em out calls} (\ocalls) to executes system calls
outside the enclave. The enclave code copies the \ocall parameters to
the untrusted partition of the application, which in turn calls the OS
system call, collects the return values, and passes it back to the
enclave. When the control returns to the enclave, the enclave wrapper
code copies the $\tt{syscall}$ return values from the untrusted memory
to the protected enclave memory. This mechanism facilitates
interactions between the enclave and non-enclave logic of an
application. A large fraction of enclave applications need to dispatch
\ocalls for standard (e.g., syscalls, libc) or application-specific
APIs.

\paragraph{Syscall Parameter Tampering.} 
This is a broad class of attacks and has been inspected in various
aspects by Ports and Garfinkel~\cite{ports-garfinkel}; a specific
subclass of it is called as Iago attacks~\cite{iago}. Ports-Garfinkel
first showed system call tampering attacks for various subsystems such
as filesystem, IPC, process management, time, randomness, and I/O. For
file content and metadata tampering attacks, their paper suggested
defenses by maintaining metadata such as a secure hash for file pages
and protecting them by MAC and freshness counter stored in the
untrusted guest filesystem. For file namespace management they
proposed using a trusted, protected daemon to maintain a secure
namespace which maps a file pathname to the associated protection
metadata. This way, checking if OS return values are correctly
computed would be easier than undertaking to compute them. An added
benefit is that the TCB of such a trusted monitoring mechanism for the
untrusted kernel is smaller. In this paper, our focus is on the
filesystem subset of the system calls. Further, we concentrate on 
enclave-like systems for Intel SGX, but our work applies equally well
to other systems~\cite{overshadow, inktag, keystone, komodo-sosp17,
sanctum}.

\input{tables/system-comparison}
\subsection{The Baseline: Existing Systems}
\label{sec:attacks}

All SGX-based systems such as Haven~\cite{haven}, Scone~\cite{scone},
\panoply~\cite{panoply}, Graphene-SGX~\cite{graphene-sgx}, Intel
Protected File System~\cite{intel-fs} which either use SDK or
hand-coded \ocall wrappers must address syscall parameter tampering
attacks. Even non-SGX TEEs have been shown to face the same
threat~\cite{keystone, overshadow, keystone-issue}. These systems are
upfront in acknowledging this gap and employ ad-hoc checks for each
API to address a subset of attacks. Using integrity preserving
filesystems~\cite{cogent} and formally testing if a filesystem abides
by POSIX semantics~\cite{sibylfs} are stepping stones towards our
goal, but they do not reason about intentional deviations by a {\em
completely Byzantine} OS. We demonstrate representative attack
capabilities on state-of-the-art enclave systems with encrypted file
storage to motivate why a provable approach down to the details is
important.

\paragraph{Baseline.}
We assume that the filesystem API uses authenticated encryption and
attestation to prevent the OS from directly tampering the file
content. Further, we assume a setting where the enclave tunnels all
file-related system / library calls to the untrusted OS. The untrusted
OS simply reads and writes encrypted blocks of data to the disk such
that the content can only be decrypted inside the enclave. Most
publicly available enclave frameworks support such a baseline defense.
For concreteness, we discuss specific details of the four open systems
available today which support a filesystem interface for enclave
applications: Graphene-SGX~\cite{graphene-sgx},
\panoply~\cite{panoply}, Intel Protected File System
Library~\cite{intel-fs}, and Google Asylo~\cite{asylo}.
Table~\ref{tbl:system-api} shows the number of file APIs supported and
the LOC of these systems indicating that custom implementations have
large TCB irrespective of the APIs they support. More importantly, as
we show in Section~\ref{sec:examples}, they employ ad-hoc checks which
do not completely defeat the attacks by the OS. As opposed to the
state-of-the-art, \codename provides provable guarantees. Our
corresponding implementation in \coq as well as hand-coded $\tt{C}$)
lowers the TCB.
\input{listings/example-app}
\input{listings/example-minipage}

\subsection{Is Encryption Sufficient?} 
\label{sec:examples}

Our baseline system encrypts and adds MAC tags to file content. We
show that this is not enough to protect against a malicious OS. We
recall attack examples from prior work and present new attacks to show
that \besfs is needed to defeat a broad class of attacks that go well
beyond memory safety.

\paragraph{Memory-safety Iago Attacks.}
Iago attacks show a subclass of concrete attacks on memory allocation
interfaces, wherein the malicious OS overlaps memory-mapped (via
$\tt{mmap}$) pages. The attack results in subverting the control flow
in the enclaved application. Iago attacks demonstrate that verifying
return values may require user-level defenses to carefully enforce
invariants on the virtual memory layout.

\paragraph{Logic Bugs via Return Value Tampering.}
We show how the OS can mislead the application-level into taking
incorrect actions, without causing a crash, by exploiting the semantic
gap between SGX guarantees and POSIX API. This attack works on
encrypted filesystems since it perpetrates by returning inconsistent
return values. Figure~\ref{lst:example-app} shows a simplified enclave
code which is executed by a node in a Tor-like service~\cite{tor}. The
enclave logic casts votes, appending it to a log file at each epoch,
say in a sub-step of a consensus process. Specifically, the enclave
function $\tt{log\_vote}$ opens an existing log file in append mode.
The enclave checks if the open was successful or were there any
errors. The function handles the error conditions and once the
$\tt{fopen}$ is successful, it writes the vote content to the file via
$\tt{fwrite}$. As per POSIX standard, the library should return a
$\tt{NULL}$ file pointer on $\tt{fopen}$ failure and set the
$\tt{errno}$ is set to indicate the error. If the file name is invalid
(e.g., empty string or a non-existing file path) the error is
$\tt{ENOENT}$. If the mode is invalid the error should be
$\tt{EINVAL}$, while $\tt{EINTR}$ indicates that the call was
interrupted and may succeed on a re-attempt. 
Figure~\ref{lst:example-app} performs error handling assuming a
POSIX-reliant filesystem.

Figure~\ref{lst:examples} shows the implementation code snippets of
file open operation in four existing SGX platforms which implement
four different types of checks. Both Graphene-SGX and \panoply simply
forward the $\tt{errno}$ to the caller without performing any checks
(Figure~\ref{lst:graphene-bug}, ~\ref{lst:panoply-bug}). In our
example (Figure~\ref{lst:example-app}), the OS can trick the enclave
into creating an empty file by falsely sending $\tt{ENOENT}$ error
code, even though the log file exists. Both the systems cannot detect
this attack. Intel's Protected File System
(Figure~\ref{lst:intel-bug}) returns an incorrect error code as per
the POSIX standards. If the enclave passes the log name to be an empty
string, the application will incorrectly receive $\tt{EINVAL}$ and
will not be able to log the vote. Google Asylo
(Figure~\ref{lst:asylo-bug}), does not perform any pre-checks on the
parameters and if the \ocall returns any errors, the system always
overwrites it with $\tt{EINTR}$ (Line $14$). Thus, our example
demonstrates that although the existing systems employ encryption on
file content, they are vulnerable to logic bugs due to incomplete
interface security checks.

\input{listings/example-glibc}
\paragraph{Glibc Logic Vulnerability due to Bad Initialization.}
We present another attack which cannot be defeated by using an
encrypted file system or sealing within the enclave. The glibc malloc
subsystem allocates large chunks of memory via anonymous $\tt{mmap}$
(Figure~\ref{lst:example-glibc} Line $13$). It then distributes and
collects parts of these chunks via $\tt{malloc}$ and $\tt{free}$
calls. For glibc's internal buffer management, the first $8$ bytes of
each mmaped region are reserved for meta-data (e.g., tracking the
sizes of the allocated chunks in Figure~\ref{lst:example-glibc} Line
$5$). The POSIX specification dictates that if the $\tt{mmap}$ syscall
requests for anonymous memory regions, which are not file-backed, the
OS must initialize the memory contents to $0$. Thus, when glibc
acquires a large buffer via anonymous-mapped memory region it assumes
that this region is filled with $0$s by the kernel. The glibc
implementation then updates the size of the current block by writing
to the size field. For the first block being mmaped, glibc does {\em
not} write $0$ to the $\tt{prev\_size}$ as it assumes those bytes are
already set to $0$.

In the implementation of $\tt{free}$, glibc unmaps chunks if all slots
in those chunks are unallocated. For this, it performs some arithmetic
computation over the start address of a chunk as well as the sizes of
the current and previous chunks. Suppose the allocated region is $[P,
P+s)$ where $P$ and $s$ denote the start address and length
respectively. Further, let $X$ denote the value of the first $8$ bytes
of a chunk i.e., variable $\tt{prev\_size}$. Lines $23$-$26$ in
Figure~\ref{lst:example-glibc} invoke the $\tt{unmap}$ syscall for the
address range $[P-X, P+s)$. In the case of the first chunk, the value
of $X$ is $0$ and glibc will unmap $[P, P+s]$ which is correct. Note
that if the OS returns mmaped memory which is filled with non-$0$s, it
can control the value of $X$. For example, if the OS selects $X \neq 0
\land X < P$, it will trick glibc into unmapping not only $[P, P+s]$
but also $[P-X, P]$. Neither glibc nor the application is aware of
this inadvertent unmapping and their internal metadata will no longer
reflect the correct state.

In general, such an attack can break the consistency enforced by
various program components. For instance, a garbage collector which
maintains invariants about how objects are traced by reference chains
may use the memory mapping information to mark the memory occupied by
freed objects to avoid use-after-free. More broadly, many security
primitives (e.g., control-flow integrity, fat pointers, taint
analyses) maintain shadow metadata at fixed offsets from program
objects, which could be affected by such inconsistency bugs.

%% file: tables/system-comparison.tex
\begin{table}[]
\centering
\resizebox{0.48\textwidth}{!}{%
\begin{tabular}{@{}lcrrrc@{}}
\toprule
\textbf{System Name} & \textbf{Release Date} & \multicolumn{1}{c}{\textbf{Total LOC}} & \multicolumn{1}{c}{\textbf{\# of APIs}} & \multicolumn{1}{c}{\textbf{FS API}} & \textbf{API Level} \\ \midrule
{Graphene-SGX} & July 2016 & 1325978 & 28 & 5 & syscall \\
{Panoply} & Dec 2016 & 20213 & 254 & 37 & libc \\
{Intel SDK} & Dec 2016 & 119234 & 24 & 15 & Custom \\
{Google Asylo} & May 2018 & 400465 & 39 & 7 & Custom \\
\textbf{BesFS} & Aug 2018 & 1449 & 21 & 13 & POSIX \\ \bottomrule
\end{tabular}%
}
\vspace{-7pt}
\caption{Comparison of existing SGX filesystem support.}
\label{tbl:system-api}
\vspace{-15pt}
\end{table}

%% file: listings/example-app.tex
\begin{figure}[t]
\centering
\begin{lstlisting}[style=JavaScript,  language=C, xleftmargin=0.3cm]
int log (char* fname, int mode, char* buf, int len) {
	int errnum, cnt = 0; FILE* fd = fopen(fname, mode);
	if (fd == NULL) {
		errnum = errno;
		if (errnum == EINVAL) fd = fopen (fname, "a"); // append
		if (errnum == ENOENT)
			if (fname == NULL) fname = "default.log";
			fd = create_log(fname); // create empty log file
		if (errnum == EINTR) fd = fopen (fname, mode); // retry
	}
	if (fd) cnt = fwrite(buf, 1, len, fd); // write log
	return cnt;
}
void cast_vote () { // each tor node ...
	status = log(log_file, mode, &vote, vote_len);
	if (status) start_election();
\end{lstlisting}
\vspace{-10pt}
\caption{
Example enclave logic. The enclave opens a log file and attempts
recovery on failure by either changing the mode (\texttt{EINVAL}), 
opening a new file since the path does not exist (\texttt{ENOENT}), or
reattempting the call (\texttt{EINTR}).}
\vspace{-15pt}
\label{lst:example-app}
\end{figure}

%% file: listings/example-minipage.tex
\begin{figure*}[t]
\centering
	\begin{subfigure}[tl]{0.49\textwidth}
	\centering
	\input{listings/graphene-bug}
	\vspace{-7pt}
	\caption{Graphene-SGX. Checks on success; otherwise forwards the error.}
	\label{lst:graphene-bug}
	\end{subfigure}
	~
	\begin{subfigure}[bl]{0.49\textwidth}
	\centering
	\vspace{-55pt}
	\input{listings/panoply-bug}
	\vspace{-5pt}
	\caption{Panoply. Forwards the fd and errors as-is if \ocall fails.}
	\label{lst:panoply-bug}
	\end{subfigure}
	\\
	\begin{subfigure}[tr]{0.49\textwidth}
	\input{listings/intel-bug}
	\caption{Intel Protected File System. Returns EINVAL instead of ENOENT.}
	\label{lst:intel-bug}
	\end{subfigure}
	~
	\begin{subfigure}[br]{0.49\textwidth}
	\centering
	\vspace{-50pt}
	\input{listings/asylo-bug}
	\caption{Google Asylo. Suppresses the error on failure and returns EINT.}
	\label{lst:asylo-bug}
	\end{subfigure}
\vspace{-8pt}	
\caption{
SGX filesystem API support. Code snippets from
four systems which support file open operation inside the enclave. 
}
\vspace{-15pt}
\label{lst:examples}	
\end{figure*}

%% file: listings/graphene-bug.tex
\begin{lstlisting}[style=JavaScript, language=C, xleftmargin=0.3cm, 
]
static int file_open (..., const char * uri, int access, int share, int create, int options) {
    int fd = ocall_open(uri, access|create|options, share);
    if (fd < 0)
        return fd;
    ...
}
static int sgx_ocall_open(void * pms) {
    ms_ocall_open_t * ms = (ms_ocall_open_t *) pms;
    int ret;
    ODEBUG(OCALL_OPEN, ms);
    ret = INLINE_SYSCALL(open, ...);
    return IS_ERR(ret)?unix_to_pal_error(ERRNO(ret)):ret;
}
\end{lstlisting}

%% file: listings/panoply-bug.tex
\begin{lstlisting}[style=JavaScript, language=C, xleftmargin=0.3cm, 
]
SGX_WRAPPER_FILE sgx_wrapper_fopen(const char* filename, const char* mode) {
	SGX_WRAPPER_FILE f = 0;
	sgx_status_t status = ocall_fopen(&f, filename, mode);
  	CHECK_STATUS(status);
	return f;
}
\end{lstlisting}

%% file: listings/intel-bug.tex
\begin{lstlisting}[style=JavaScript, language=C, xleftmargin=0.3cm, 
]
SGX_FILE* sgx_fopen
(const char* filename, const char* mode) {
	return sgx_fopen_internal(filename, mode, NULL, key);
}
static SGX_FILE* sgx_fopen_internal
(const char* filename, const char* mode) {
	protected_fs_file* file = NULL;
	if (filename == NULL || mode == NULL)	{
		errno = EINVAL;
		return NULL;
	}
    ...
}
\end{lstlisting}

%% file: listings/asylo-bug.tex
\begin{lstlisting}[style=JavaScript, language=C, xleftmargin=0.3cm, 
]
int secure_open(const char *pathname, int flags, ..){
	...
	bool is_new_file = (enc_untrusted_access(pathname, F_OK) == -1);
	int fd = enc_untrusted_open(pathname, flags, mode);
	if (fd == -1) 
		return -1;
	...
}
int enc_untrusted_open(const char *path_name, int flags) {
	uint32_t mode = 0;
	int result;
	sgx_status_t status = ocall_enc_untrusted_open(&result, path_name, flags, mode);
	if (status != SGX_SUCCESS) {
		errno = EINTR;
		return -1;
	}
	return result;
}
\end{lstlisting}

%% file: listings/example-glibc.tex
\begin{figure}[]
\centering
\begin{lstlisting}[style=JavaScript,  language=C, xleftmargin=0.3cm]
/* There is only one instance of the malloc params. */
static struct malloc_par mp_ = {...};
typedef struct malloc_chunk {
    size_t  prev_size;  /* previous chunk size(if free) */
    size_t  size;       /* Size (bytes) including metadata */
    ...
}mchunkptr;
static void *sysmalloc (INTERNAL_SIZE_T nb, mstate av) {
    ...
    mchunkptr p;        /* the allocated/returned chunk */
    char *mm;           /* return value from mmap call */
    ...
    mm = (char *) (mmap (0, size, PROT_READ | PROT_WRITE, 0));
    ...
    p = (mchunkptr) mm;
    p->size = size | IS_MMAPPED;
    ....
    return chunk2mem (p);
    ...
}
static void munmap_chunk (mchunkptr p) {
    ...
    uintptr_t block = (uintptr_t) p - p->prev_size;
    size_t total_size = p->prev_size + size;
    ...
    munmap ((char *) block, total_size);
    ...
}
\end{lstlisting}
\vspace{-5pt}
\caption{Glibc Attack. OS corrupts \texttt{prev\_size} via
\texttt{mmap} (Line $13$). It can trick glibc into inadvertently
unmapping larger memory range (Line $26$) without the updating 
glibc's internal metadata which violates its constraints.
}
\vspace{-15pt}
\label{lst:example-glibc}
\end{figure}

%% file: chapters/design.tex
\section{\codename Design}
\label{sec:formulation}

All the classes of filesystem API attacks presented in
Section~\ref{sec:examples} stem from the fact that the OS can deviate
from its expected semantics. Our goal is to design a filesystem
interface, called \codename, which protects the enclave from a broad
category of such attacks. These attacks include (but are not limited
to) Iago attacks, file content manipulation such as mapping multiple
file blocks of the same or different files to single physical block,
operating on content at the wrong offset or block, and misaligned
sequences of file blocks in a file. Further, the OS can perpetrate
mismatch attacks by ignoring the user-provided parameters such as
paths, file descriptor, or size (e.g., violate the size requested in
the operations). Lastly, it can change the error codes returned by the
filesystem and force the enclave to execute a different control-flow
path.

\input{chapters/approach}

\input{chapters/interface}

%% file: chapters/approach.tex
\subsection{Approach}
\label{sec:approach}
We seek for the right abstraction which is necessary to capture the
filesystem behavior inside the enclave as well as sufficient to detect
any deviation from the Byzantine OS. Attacks on an enclave can arise
at multiple layers of the filesystem stack. Our choice of the layer
where we formally proof-check the \codename API is guided by the
observation that the higher the layer we safeguard, the {\em larger}
the attack surface (i.e., TCB) we can eliminate, and the more
implementation-agnostic the \codename API becomes. One could include
all the layers starting at the disk kernel driver, where content is
finally mapped to persistent storage, in the enclave TCB. Enforcing
safety at this interface will require simply encrypting/decrypting
disk blocks with correct handling for block positions~\cite{sego}.
Alternatively, one could include a virtual filesystem management
layer, which maps file abstractions to disk blocks and physical page
allocations, in the enclave---as done in several LibraryOS systems
like Graphene-SGX~\cite{graphene-sgx,haven}. To ensure safety at this
layer, the model needs to reason about simple operations (reads,
writes, sync, and metadata management). Further up, one could design
to protect at the system call layer, leaving all of the logic for a
filesystem (e.g., journaling, physical page management, user
management, and so on) outside the enclave TCB. However, this still
includes the entire library code (e.g., the \libc logic) which manages
virtual memory of the user-level process (heap management, allocation
of user-level pages to buffers, and file-backed pages). For instance,
this is $1.29$ MLOC and $88$ KLOC in glibc and musl-libc,
respectively. Once we include such a TCB inside the enclave, we either
need to prove its implementation safety or trust it with blind faith.
We decide to model our API above these layers, excluding them from the
TCB.

\codename models the POSIX standard for file sub-systems. POSIX is a
documented standard, with small variations across implementations on
various OSes~\cite{sibylfs}. In contrast, many of the other layers do
not have such defined and stable interfaces. At the POSIX layer,
\codename models the file / directory path structures, file content
layouts, access rights, state metadata (file handles, position
cursors, and so on). Specifically, \codename ensures safety without
the need to model virtual-to-physical memory management, storage,
specifics of kernel data structures for namespace management (e.g.,
Linux inode, user groups), and so on. \codename is thus generic and
compatible with different underlying filesystem implementations (NFS,
ext4, and so on). Further, this API choice reduces the proof
complexity as they are dispatched for simpler data structures.

\paragraph{Solution Overview.} \codename is an abstract filesystem
interface which ensures that the OS follows the semantics of a benign
filesystem---it is exhibiting observationally equivalent behavior to a
good OS. This way, instead of enlisting potentially an infinite set of
attacks, we define a good OS and deviation from it is categorized as
an attack from a compromised or a potentially malicious OS.
Specifically, our definition of a good OS not only includes
POSIX-compliance but also a set of safety properties expected from the
underlying filesystem implementation. We design a set of $15$ core
filesystem APIs along with a safety specification. Table~\ref{tab:api}
shows this \codename POSIX-compliant interface, which can be invoked
by an external client program running in the SGX enclave. It has a set
of {\em methods}, {\em states}, and {\em safety properties}
(\SP{1}-\SP{5} and \TP{1}-\TP{15}) defined in
Section~\ref{sec:interface}. Each method operates on a starting state
(implicitly) and client program inputs. The safety properties capture
our definition of a benign OS behavior. Empirically, we show in
Section~\ref{sec:eval} that the real implementations of existing OS,
when benign, satisfy the \codename safety properties---the application
executes with the \codename interface as it does with direct calls to
the OS. Further, the safety properties reject any deviations from a
benign behavior, which includes all the above attacks. Thus, \codename
is a state transition system. We define a good start state that
satisfies the state properties (\SP{1}-\SP{5}). Our transition
properties (\TP{1}-\TP{15}) ensure that the file system is in a good
state after executing a \codename API call. 

Importantly, we prove that the safety of \codename API is {\em
serially composable}. This composability is crucial to allow
executions of benign applications that make a potentially infinite set
of calls. Further, one can model higher-level API (e.g., the
$\tt{fprintf}$ interface in $\tt{libc}$) by composing two or more
\codename API operations. Thus, composition property allows us to
reduce the size of the core APIs that have to be proved as well as
reduce the attack surface for the OS. To ensure serial composition,
the state safety properties (\SP{1}-\SP{5}) enforce that if we invoke
a \codename core API operation in a good (safe) state, we are
guaranteed to resume control in the application in a good state.
Second, we show that calls can be chained, i.e., the good state after
a call can be used as an input to any of the \codename calls, through
a set of safe transition properties (\TP{1}-\TP{15}). We provide a
machine-checked \coq implementation of the \codename API
(Section~\ref{sec:interface}).

\begin{theorem}[State Transition Safety.]\label{thm:safety}
Given a good state $\stateorig$ satisfying ${\text{pre}}_i$, if we
execute $f_i$ to reach state $\stateprime$, then $\stateprime$ is
always a good state and the relation between $\stateorig$ and
$\stateprime$ is valid according to the transition relation $\tau_i$:
\begin{gather*}
\forall \stateorig, \stateprime,i. ~~~ \stateorig \models \text{\SP{1}--\SP{5}} ~ \wedge ~ 
{\text{pre}}_i(\stateorig) ~ \wedge ~ \stateorig \stackrel{f_i} 
{\leadsto} \stateprime ~~~ \Rightarrow \null \\
\tau_i(\stateorig, \stateprime) ~ \wedge ~ \stateprime \models {\text{\SP{1}--\SP{5}}}
\end{gather*}
\end{theorem}
We can verify sequences of calls to our API by inductively
chaining this theorem. Our second theorem states that the state
property is preserved for a composition of any sequence of interface
calls. We close the proof loop with induction by starting in a good
initial state and using Theorem~\ref{thm:safety} to show that a method invocation
in \codename always produces a good state for a sequential composition
of transitions. \coq proof assistant dispatches the proof.
\begin{theorem}[Sequential Composition Safety.]\label{thm:composition}
Given a good initial state $\stateorig_{\text{0}}$ subject to a
sequence of transitions $\tau_{m_1}, \dots, \tau_{m_n}$ always
produces a good final state $\stateorig_n$:
\begin{gather*}
\stateorig_{\text{0}} \models \text{\SP{1}--\SP{5}} \; \wedge 
\stateorig_0 \stackrel{f_{m_1}}{\leadsto} \stateorig_1 \; \wedge
\stateorig_1 \stackrel{f_{m_2}}{\leadsto} \stateorig_2 \; \wedge \dots \wedge
\stateorig_n \stackrel{f_{m_n}}{\leadsto} \stateorig_n \; \\ \Rightarrow \wedge
\tau_{m_1} (\stateorig_0, \stateorig_1) \; \wedge \;
\tau_{m_2} (\stateorig_1, \stateorig_2) \; \wedge \dots \; \wedge 
\tau_{m_n} (\stateorig_{n - 1}, \stateorig_n) \; \wedge\; \\
\stateorig_n \models \text{\SP{1}--\SP{5}}
\end{gather*}
\end{theorem}

\paragraph{Scope.}
We limit the scope of \codename goals in two ways:
\begin{itemize}
\squish	

\item 
For safety and simplicity, \codename filesystem state and API
intentionally does not include all the features in a typical
full-fledged filesystem. The enclave files can be concurrently
accessed by non-enclave applications, as long as the applications
abide by the safety restrictions enforced by \codename. We detect if
any entity (other enclaves, user applications, the OS) violates
\codename invariants and abort the enclave. 

\item 
\codename aims strictly at integrity property. Several known
side-channels and hardware mistakes impact the confidentiality
guarantees of SGX~\cite{cca-sgx, spectre}. Out of the $167$ lemmas in
\codename, only one lemma assumes the correctness of the cryptographic
operations. Specifically, \codename assumes the secrecy of its AES-GCM
key used to ensure the integrity of the filesystem content. Our lemma
assumes that the underlying cryptography does not allow the adversary
to bypass the integrity checks by generating valid tags for arbitrary
messages. Further, we assume that the adversary does not know the
AES-GCM key used by the enclave to generate the integrity tags.
Higher-level confidentiality guarantees are not within the scope of
\codename goals (c.f.~\cite{slashconfidential, moat,
sgx-policies-oopsla16}).

\end{itemize}

%% file: chapters/interface.tex
\subsection{\codename Interface}
\label{sec:interface}
\codename interface is a state transition system. It defines a set of
valid filesystem states and methods to move from one state to another.
While doing so, \codename dictates which transitions are valid by a
set of transition properties.

\paragraph{State.}
\codename has type variables which together define a state. We choose
to include minimal filesystem metadata in the \codename state while
providing maximum expressiveness in its APIs. This selection is
inspired by our survey of previous filesystem verification efforts for
various purposes~\cite{fs-veri-too, cogent, sibylfs}. Specifically,
\codename state comprises valid paths in the filesystem (\pathset),
mappings of paths to file / directory identifiers and metadata
(\nodemap), set of open files (\oh), memory maps of file content
(\memory), memory map of anonymously mmaped page content (\ammap), and
anonymous page mapping metadata (\mh). We define them as follows:

\input{chapters/types}

\paragraph{State Properties.}
The state variables cannot take arbitrary values. They must abide by a
set of state properties defined by \codename stated in
Table~\ref{table:sp}. For path set $\pathsetm$, \codename enforces
that the entries in the path set are unique and do not contain
circular paths. This ensures that each directory contains unique file
and directory names by the definition of a path set. All files and
directories in \codename have unique identifiers and are mapped by the
partial function \nodemap to their metadata such as permission bits
and size, stated formally as \SP{1}. All open file IDs have to be
registered in the \oh (\SP{2}). \oh can only have unique entries
(\SP{3}) and the cursor of an open file handle cannot take a value
larger than that file's current size (\SP{4}). As per \SP{5}, \memory
and \ammap do not allow any overlaps between addresses and have a
one-to-one mapping from the virtual address to content. The partial
functions for \memory and \ammap ensures this by definition. All file
operations are bounded by the file size and all anonymous memory
dereferences are bounded by the size of the allocated memory. 
Specifically, the file memory can be dereferenced only for offsets
between $0$ and the $\tt{EOF}$. Any attempts to access file content
beyond $\tt{EOF}$ are invalid by definition in \codename and is
represented by the symbol $\bot$. Similarly, the current cursor
position can only take values between $0$ and $\tt{EOF}$ (\SP{5}).
\input{tables/state-property}

\input{tables/safety-table}

\paragraph{Transition Properties.}
\codename interface specifies a set of methods listed in \codename API
in Table~\ref{tab:api}. Each of these methods takes in a valid state
and user inputs to transition the filesystem to a new state. 
\codename interface facilitates safe state transitions. Formally, we
represent it as $\tau_{m_i}(\stateorig, \stateprime,
\overrightarrow{\text{out}})$, where $\tau_{m_i}$ is the interface
method invoked on state $\stateorig$ to produce a new state
$\stateprime$. The vector $\overrightarrow{\text{out}}$ represents the
explicit results of the interface. This way, \codename enforces {\em
state transition atomicity} i.e., if the operation is completed
successfully then all the changes to the filesystem must be reflected;
if the operation fails, then \codename does not reflect any change to
the filesystem state.

\paragraph{\codename Safety Guarantees.}
\codename satisfies the state properties at initialization because the
start state ($\stateorig_{\mathtt{init}}$) is empty. Specifically, all
the lists are empty and the mappings do not have any entries. So, they
trivially abide by the state properties in
($\stateorig_{\mathtt{init}}$). Once the user starts interfacing with
the \codename state, we ensure that \codename state properties
(\SP{1}-\SP{5}) still hold. Further, each interface itself dictates a
set of constraints (e.g., the file should be opened first to close
it). Thus, interface-specific properties not only ensure that the
state is valid but also specify the safe behavior for each interface.
Transition properties \TP{1}-\TP{15} (Table~\ref{tab:api}) define type
map, state, and state transition for \codename interface.

\subsection{How Do Our Properties Defeat Attacks?}

Our state properties in Section~\ref{sec:interface} and transition
properties in Table~\ref{tab:api} are strong enough to defeat the OS
attacks.

\paragraph{File \& Memory Content Manipulation (A1).}
Our baseline encrypts all the file data blocks and anonymously mmapped
content which prevents direct tampering from the OS. However, there
are other avenues of attacks beyond this which \codename captures.
Specifically, the unique mapping property (\SP{5}) of \memory and
\ammap ensures that the OS cannot go undetected if it reorders or
overlaps the underlying pages of the file content or anonymous mmaps.

\paragraph{Path Mismatch (A2a).}
\codename state ensures that each path is uniquely mapped to a file or
directory node. All methods which operate on paths first check if the
path exists and if the operation is allowed on that file or directory
path. For example, for a method call $\tt{readdir(``foo/bar")}$, the
path $\tt{foo/bar}$ may not exist or can be a file path instead of a
directory path. \SP{1} ensures that file directory paths are distinct,
unique, and mapped to the right metadata information. Subsequently,
any queries or changes to the path structure ensure that these
properties are preserved. For example, $\tt{fs\_create}$ checks if the
parent path is valid and if the file name pre-exists in the parent
path. The corresponding state is updated if all the pre-conditions are
met (\SP{4}). 

\paragraph{File Descriptor Mismatch (A2b).}
Once the file is opened successfully, all file-content related
operations are facilitated via the file descriptor. \codename ensures
that the mappings from the file name to the descriptor are unique and
are preserved while the file is open. Further, \codename maps any
updates to the metadata or file content via the file descriptor such
that it detects any mapping corruption attempts from the OS (\SP{5}).

\paragraph{Size Mismatch (A3).}
\codename's atomicity property ensures that the filesystem completely
reflects the semantics of the interface during the state transition.
Our file operations have properties which ensure that \codename
operates on the size specified in the input. $\tt{fs\_read}$,
$\tt{fs\_write}$, and $\tt{fs\_truncate}$ post-conditions reflect this
in Table~\ref{tab:api}.

\paragraph{Error Code Manipulation (A4).}
All state or transition property violations in the interface execution
map to a specific error code. Each of these error codes distinctly
represents which property was violated. For example, if the user tries
to read using an invalid file descriptor, the \SP{3} and \TP{11}
properties are violated and \codename return an $\tt{eBadF}$ error
code. If there are no violations and the state transition succeeds,
\codename returns the new filesystem state and $\eSucc$. \codename
interface performs its own checks to identify error states. This way,
we ensure that the OS cannot go undetected if it attempts to
manipulate the enclave with wrong error codes.

\paragraph{Iago \& Libc Attacks.}
\codename defends against a broader class of attacks, including Iago
attacks, because we check all the return values after a file-related
system call. We ensure that the values are correct by checking it
against the in-enclave state of the filesystem. For anonymous mmap,
\codename checks if the untrusted memory region returned by the OS is
indeed zeroed out. \codename makes a copy of the mmaped memory inside
the enclave and all accesses to the mmaped memory are redirected to
the in-enclave address.

%% file: chapters/types.tex
{\footnotesize
\begin{align*}
\pathsetm \coloneqq \{\mathtt{p} \mid \mathtt{p}: \mathtt{Path} \} \; \;\; \; &
\nodemapm \coloneqq \mathtt{Path} \nrightarrow 
\mathtt{Id} \cross \mathtt{Permission}\cross\mathtt{Size} \\
\ammapm \coloneqq \mathbb{N}\nrightarrow \mathtt{Byte} \; \; \; \; \; &
\mhm \coloneqq \{(\mathtt{sAddr}, \mathtt{length}) \mid \mathtt{sAddr}: 
\mathbb{N}, \mathtt{length}:\mathbb{N}\} \\
\memorym \coloneqq \mathtt{Id}\cross\mathbb{N}\nrightarrow \mathtt{Byte} \; \; \; &
\ohm \coloneqq \{(\mathtt{fileId}, \mathtt{cursor}) \mid 
\mathtt{fileId}: \mathtt{Id}, \mathtt{cursor}:\mathbb{N} \} 
\end{align*}
}%
All file and directory paths in the filesystem are captured by path
set $\pathsetm$, where $\mathtt{Path}$ represents the path data type.
A directory path type is denoted by $\pathsetm_{\textsc{Dir}}$,
whereas a file path type is denoted by $\pathsetm_{\textsc{File}}$. We
define the $\mathsf{Parent}$ operator which takes in a path and
returns the parent path. For example, if the path $p$ is
$\tt{/foo/bar/file.txt}$, then $\mathsf{Parent}(p)$ gives the parent
path $\tt{/foo/bar}$. \codename captures the information about the
files and directories via the node map $\nodemapm$. \codename
allocates an identifier to each file and directory for simplifying the
operations which operate on file handles instead of paths. We
represent the user read, write, and execute permissions by
$\tt{Permission}$. The size field for a file signifies the number of
bytes of file content. For directories, the size is supposed to
signify the number of files and directories in it. For simplicity,
\codename currently does not track the number of elements in the
directory and all the size fields for all the directories are always
set to $0$. For a path $p$, we use the subscript notations
$\nodemapm(p)_{\mathtt{Name}}$, $\nodemapm(p)_{\mathtt{Id}}$,
$\nodemapm(p)_{\mathtt{perm}}$, and $\nodemapm(p)_{\mathtt{Size}}$ to
denote the name, id, permissions, and size respectively. Each open
file is tracked using $\ohm$ via its file id. $\ohm$ tracks the
current cursor position for the open file to facilitate operations on
the file content. Given a tuple $o \in \ohm$, for simplicity, we use
subscript notations $o_{\mathtt{Id}}$ and $o_{\mathtt{Cur}}$ to denote
the id and the cursor position of that file. The file content is
stored in a byte memory and each byte can be accessed using the tuple
comprising file id and a specific position. The anonymously mapped
memory is stored in a separate byte memory and can be accessed using a
position. $\mhm$ tracks the anonymous memory allocations which
include the start position and total length of each mapping. Thus,
\codename state $\stateorig_\text{\codename}$ is defined by the tuple
$\langle \pathsetm, \nodemapm, \mhm, \ammapm, \ohm, \memorym \rangle$.
Note that the \besfs API includes calls to open and close the
filesystem. The user can use these calls to persist the internal state
of \besfs  inside the enclave for reboots and crash recovery similar
to traditional filesystems~\cite{fscq}. More importantly, these two
APIs ensure that the filesystem has temporal integrity to prevent
rollbacks. \codename ensures that the enclave sees
the last saved state on reboot/restart.

%% file: tables/state-property.tex
\begin{table}[t]
\resizebox{0.48\textwidth}{!}{%
\begin{tabular}{@{}cc@{}}
\toprule
{\textbf{$\textsf{SP}_{i}$}} & \textbf{State Property Definition} \\ \midrule
{$\textsf{SP}_{1}$} & $\mathrm{dom}(\nodemapm)=\pathsetm \forall (p, p') \in \pathsetm\cross\pathsetm, p \neq p' \Rightarrow \nodemapm(p)_{\mathtt{Id}} \neq \nodemapm(p')_{\mathtt{Id}}$ \\
{$\textsf{SP}_{2}$} & $\forall o\in \ohm, \exists p\text{~s.t.~} p\in\pathsetm \wedge \nodemapm(p)_{\mathtt{Id}}=o_{\mathtt{Id}}$ \\ 
{$\textsf{SP}_{3}$} & $\forall (o, o') \in \ohm \cross\ohm , o_{\mathtt{Id}} = o'_{\mathtt{Id}} \Rightarrow o = o'$ \\ 
{$\textsf{SP}_{4}$} & $\forall p\in\pathsetm, o\in\ohm, \nodemapm(p)_{\mathtt{Id}}=o_{\mathtt{Id}} \; \Rightarrow \; o_{\mathtt{Cursor}}<\nodemapm(p)_{\mathtt{Size}}$ \\ 
{$\textsf{SP}_{5}$} & $\forall \mathtt{f}, \forall \mathtt{o}, \exists p \mathtt{~s.t.~} p \in \pathsetm \, \wedge \, \mathtt{f}=\nodemapm(p)_{\mathtt{Id}} \, \wedge \, \mathtt{o} < \nodemapm(p)_{\mathtt{Size}} \Rightarrow \memorym(f,o) \neq \bot$ \\ \bottomrule
\end{tabular}
}
\caption{
\codename State Properties. Formal definitions of the state
properties enforced at any point in time.
}
\vspace{-15pt}
\label{table:sp}
\end{table}

%% file: tables/safety-table.tex
\begin{table*}[t]
\centering
\resizebox{\textwidth}{!}{%
\begin{tabular}{@{}lrlclll@{}}%
\toprule
\multicolumn{1}{c}{\textbf{$\textsf{TP}_{i}$}} & \multicolumn{2}{c}{\textbf{\codename Interface}} & \textbf{Pre-condition $\, \textsf{Pre}_{i} (\stateorig)$} & \multicolumn{3}{c}{\textbf{Transition Relation $\, {\tau}_{i}(\stateorig, \stateprime)$}} \\ \midrule
\multirow{2}{*}{$\textsf{TP}_{1}$} & fs\_close & $(h:\mathtt{Id})$ & \multirow{2}{*}{$\exists o, \, o_{\mathtt{Id}} = h \, \wedge \, o \in \ohm$} & \multirow{2}{*}{$\stateprime \, = \, \stateorig [\ohm/\ohm-\{o\}]$} & \multirow{2}{*}{$\wedge$} & \multirow{2}{*}{$e = \eSucc $} \\
 & $\produces$ & $(e:\mathtt{Error})$ & & & & \\ \hline
\multirow{3}{*}{$\textsf{TP}_{2}$} & fs\_open & $(p:\mathtt{Path})$ & $p\in\pathsetm \, \wedge$ & \multirow{3}{*}{$\stateprime\, = \,\stateorig [\ohm/ \ohm + \{(\nodemapm(p)_{\mathtt{Id}}, \, 0)\}]$} & \multirow{3}{*}{$\wedge$} & $e = \eSucc \, \wedge$ \\
 & \multirow{2}{*}{$\produces$} & $(h:\mathtt{Id},$ & \multirow{2}{*}{$\forall o\in\ohm, \, \nodemapm(p)_{\mathtt{Id}} \neq o_{\mathtt{Id}}$} & & & \multirow{2}{*}{$h = \nodemapm(p)_{\mathtt{Id}}$} \\
 & & $e:\mathtt{Error})$ & & & & \\ \hline
\multirow{3}{*}{$\textsf{TP}_{3}$} & \multirow{2}{*}{fs\_mkdir} & $(p:\mathtt{Path},$ & $p\notin\pathsetm \, \wedge \, \mathsf{Parent}(p) \in \pathsetm_{\textsc{Dir}} \, \wedge \,$ & \multirow{2}{*}{$\stateprime\, = \,\stateorig [\pathsetm / \pathsetm + \{p\},$} & \multirow{3}{*}{$\wedge$} & \multirow{3}{*}{$e = \eSucc$} \\
 & & $r:\mathtt{Perm})$ & \multirow{2}{*}{$\nodemapm(\mathsf{Parent}(p))_{\textsc{W}}=\true$} & & & \\
 & $\produces$ & $(e:\mathtt{Error})$ & & $\qquad \quad \nodemapm / \nodemapm \bigoplus (p \mapsto \langle h, \, r, \, 0 \rangle)]$ & & \\ \hline
\multirow{3}{*}{$\textsf{TP}_{4}$} & \multirow{2}{*}{fs\_create} & $(p:\mathtt{Path},$ & $p\notin\pathsetm \, \wedge \, \mathsf{Parent}(p) \in \pathsetm_{\textsc{Dir}} \, \wedge$ & $\stateprime\, = \,\stateorig [\pathsetm / \pathsetm + \{p\},$ & \multirow{3}{*}{$\wedge$} & \multirow{3}{*}{$e = \eSucc$} \\
 & & $r:\mathtt{Perm})$ & \multirow{2}{*}{$\nodemapm(\mathsf{Parent}(p))_{\textsc{W}}=\true$} & \multirow{2}{*}{$\qquad \quad \nodemapm / \nodemapm\bigoplus (p \mapsto \langle h, \, r, \, 0 \rangle)]$} & & \\
 & $\produces$ & $(e:\mathtt{Error})$ & & & & \\ \hline
\multirow{3}{*}{$\textsf{TP}_{5}$} & fs\_remove & $(p:\mathtt{Path})$ & $p\in\pathsetm_{\textsc{File}} \, \wedge$ & \multirow{3}{*}{$\stateprime\, = \,\stateorig [\pathsetm / \pathsetm - \{p\}]$} & \multirow{3}{*}{$\wedge$} & \multirow{3}{*}{$e = \eSucc$} \\
 & \multirow{2}{*}{$\produces$} & \multirow{2}{*}{$(e:\mathtt{Error})$} & \multirow{2}{*}{$\nodemapm(\mathsf{Parent}(p))_{\textsc{W}}=\true$} & & & \\
 & & & & & & \\ \hline
\multirow{3}{*}{$\textsf{TP}_{6}$} & fs\_rmdir & $(p:\mathtt{Path})$ & $p\in\pathsetm_{\textsc{Dir}} \, \wedge \, \forall q \in \pathsetm, \, \mathsf{Parent}(q) \neq p \, \wedge$ & \multirow{3}{*}{$\stateprime\, = \,\stateorig [\pathsetm / \pathsetm - \{p\}]$} & \multirow{3}{*}{$\wedge$} & \multirow{3}{*}{$e = \eSucc$} \\
 & \multirow{2}{*}{$\produces$} & \multirow{2}{*}{$(e:\mathtt{Error})$} & \multirow{2}{*}{$\nodemapm(\mathsf{Parent}(p))_{\textsc{W}}=\true$} & & & \\
 & & & & & & \\ \hline
\multirow{4}{*}{$\textsf{TP}_{7}$} & fs\_stat & $(h:\mathtt{Id})$ & \multirow{2}{*}{$\exists o,\,o_{\mathtt{Id}} = h\, \wedge \, o \in \ohm \, \wedge$} & \multirow{4}{*}{$\stateprime\, = \,\stateorig$} & \multirow{4}{*}{$\wedge$} & $e = \eSucc \, \wedge$ \\
 & \multirow{3}{*}{$\produces$} & $(r:\mathtt{Perm},$ & & & & $r = \nodemapm(p)_{\mathtt{Perm}}\, \wedge$ \\
 & & $n: \mathtt{String},$ & \multirow{2}{*}{$\exists p, \nodemapm(p)_{\mathtt{Id}} = h \, \wedge \, p \in \pathsetm_{\textsc{File}}$} & & & $l = \nodemapm(p)_{\mathtt{Size}} \, \wedge$ \\
 & & $l: \mathbb{N}, \, e: \mathtt{Error})$ & & & & $ n = \nodemapm(p)_{\mathtt{Name}}$ \\ \hline
\multirow{3}{*}{$\textsf{TP}_{8}$} & fs\_readdir & $(p:\mathtt{Path})$ & \multirow{3}{*}{$p\in\pathsetm_{\textsc{Dir}}$} & \multirow{3}{*}{$\stateprime\, = \,\stateorig$} & \multirow{3}{*}{$\wedge$} & $e = \eSucc \, \wedge$ \\
 & \multirow{2}{*}{$\produces$} & ($l: [String],$ & & & & \multirow{2}{*}{$\forall n \in l, \, p + n \in \pathsetm$} \\
 & & $e: \mathtt{Error})$ & & & & \\ \hline
\multirow{3}{*}{$\textsf{TP}_{9}$} & \multirow{2}{*}{fs\_chmod} & $(p:\mathtt{Path},$ & \multirow{3}{*}{$p\in\pathsetm$} & \multirow{3}{*}{$\stateprime\, = \,\stateorig [\nodemapm / \nodemapm \bigodot (p \mapsto \langle\nodemapm(p)_{\mathtt{Id}}, \,r, \, \nodemapm(p)_{\mathtt{size}}\rangle)]$} & \multirow{3}{*}{$\wedge$} & \multirow{3}{*}{$e = \eSucc$} \\
 & & $ \, r:\mathtt{Perm})$ & & & & \\
 & $\produces$ & $(e:\mathtt{Error})$ & & & & \\ \hline
\multirow{3}{*}{$\textsf{TP}_{10}$} & \multirow{2}{*}{fs\_seek} & $(h:\mathtt{Id},$ & \multirow{2}{*}{$\exists o,\,o_{\mathtt{Id}} = h\, \wedge \, o \in \ohm \, \wedge$} & \multirow{3}{*}{$\stateprime \,= \,\stateorig [\ohm/\ohm-\{o\} + \{(h, l)\}]$} & \multirow{3}{*}{$\wedge$} & \multirow{3}{*}{$e = \eSucc$} \\
 & & $l : \mathbb{N})$ & & & & \\
 & $\produces$ & $(e:\mathtt{Error})$ & $\exists p, \nodemapm(p)_{\mathtt{Id}} = h \, \wedge \, l \,< \,\nodemapm(p)_{\mathtt{Size}}$ & & & \\ \hline
\multirow{4}{*}{$\textsf{TP}_{11}$} & \multirow{2}{*}{fs\_read} & $(h:\mathtt{Id}, $ & \multirow{2}{*}{$\exists o,\,o_{\mathtt{Id}} = h\, \wedge \, o \in \ohm \, \wedge$} & \multirow{4}{*}{$\stateprime \,= \,\stateorig [\ohm/\ohm-\{o\} + \{(h, o_{\mathtt{Cur}}+ l)\}]$} & \multirow{4}{*}{$\wedge$} & \multirow{2}{*}{$e = \eSucc \, \wedge$} \\
 & & $l : \mathbb{N})$ & & & & \\
 & \multirow{2}{*}{$\produces$} & $(b: [\mathtt{Byte}],$ & \multirow{2}{*}{$\exists p, \nodemapm(p)_{\mathtt{Id}} = h \, \wedge \, o_{\mathtt{Cur}} + l \, < \, \nodemapm(p)_{\mathtt{Size}}$} & & & \multirow{2}{*}{\begin{tabular}[c]{@{}l@{}}$b = \memorym (h, o_{\mathtt{Cur}}), \dots,$\\ \qquad$\memorym (h,o_{\mathtt{Cur}}+ l))$\end{tabular}} \\
 & & $e:\mathtt{Error})$ & & & & \\ \hline
\multirow{4}{*}{$\textsf{TP}_{12}$} & \multirow{3}{*}{fs\_write} & $(h:\mathtt{Id},$ & \multirow{2}{*}{$\exists o,\,o_{\mathtt{Id}} = h\, \wedge \, o \in \ohm \, \wedge$} & \multirow{2}{*}{$\stateprime \,= \,\stateorig [\ohm/\ohm-\{o\} + \{(h, l + b_{\mathtt{len}})\},$} & \multirow{4}{*}{$\wedge$} & \multirow{4}{*}{$e = \eSucc$} \\
 & & $l : \mathbb{N}, $ & & & & \\
 & & $b: [\mathtt{Byte}])$ & \multicolumn{1}{l}{\multirow{2}{*}{$\exists p, \nodemapm(p)_{\mathtt{Id}} = h \, \wedge \, l \, < \, \nodemapm(p)_{\mathtt{Size}}$}} & \multirow{2}{*}{\begin{tabular}[c]{@{}l@{}}$\qquad\quad\memorym / \memorym \bigodot ( (h, l) \mapsto b[0] , \dots ,$ \\ \qquad \qquad \qquad \quad $\; ((h, l+b_{\mathtt{len}}) \mapsto b[b_{\mathtt{len}}])]$\end{tabular}} & & \\
 & $\produces$ & $(e:\mathtt{Error})$ & \multicolumn{1}{l}{} & & & \\ \hline
\multirow{3}{*}{$\textsf{TP}_{13}$} & \multirow{2}{*}{fs\_truncate} & $(h:\mathtt{Id},$ & \multirow{2}{*}{$\exists o,\,o_{\mathtt{Id}} = h\, \wedge \, o \in \ohm \, \wedge$} & \multirow{3}{*}{$\stateprime\, = \,\stateorig [\nodemapm / \nodemapm \bigodot (p \mapsto \langle \nodemapm(p)_{\mathtt{Id}}, \, \nodemapm(p)_{\mathtt{perm}}, \, l \rangle)]$} & \multirow{3}{*}{$\wedge$} & \multirow{3}{*}{$e = \eSucc$} \\
 & & $ l : \mathbb{N})$ & & & & \\
 & $\produces$ & $(e:\mathtt{Error})$ & $\exists p, \nodemapm(p)_{\mathtt{Id}} = h \, \wedge \, l \, < \, \nodemapm(p)_{\mathtt{Size}}$ & & & \\ \hline
\multirow{3}{*}{$\textsf{TP}_{14}$} & \multirow{2}{*}{fs\_mmap} & $(l:\mathbb{N})$ & \multirow{2}{*}{$ l > 0$ } & \multirow{2}{*}{$\stateprime\, = \,\stateorig [\mhm / \mhm + \{(a ,\, l)\}, \ammapm / \ammapm \bigodot ([a] \mapsto 0, \dots, [a+l-1] \mapsto 0) ]$} & \multirow{3}{*}{$\wedge$} & \multirow{3}{*}{$e = \eSucc$} \\
 & $\produces$ & $(a:\mathbb{N},e:\mathtt{Error})$ & & & & \\ \hline
\multirow{3}{*}{$\textsf{TP}_{15}$} & \multirow{2}{*}{fs\_unmmap} & $(a:\mathbb{N})$ & \multirow{2}{*}{$ \exists q, q_{\mathtt{sAddr}} = a \land q \in \mhm$ } & \multirow{3}{*}{$\stateprime\, = \,\stateorig [\mhm / \mhm - \{(a ,\, q_{\mathtt{length}})\}] $} & \multirow{3}{*}{$\wedge$} & \multirow{3}{*}{$e = \eSucc$} \\
& $\produces$ & $(e:\mathtt{Error})$ & & & & \\ \bottomrule 

\end{tabular}
}
\caption{
\codename Interface. Method API, pre-conditions, transition relations
and post-conditions. $\stateprime = \stateorig[\mathcal{K}/
\mathcal{{K}^{\prime}}]$ denotes everything in $\stateprime$ is the
same as $\stateorig$, only $\mathcal{K}$ is replaced with
$\mathcal{{K}^{\prime}}$. In Column $4$, the $-$ and $+$ symbols
denote set addition and deletion operations. $\bigoplus$ denotes new
mapping is added and $\bigodot$ denotes update of a mapping in
relation.
}
\vspace{-10pt}
\label{tab:api}
\end{table*}

%% file: chapters/impl.tex
\section{\codename Implementation}
\label{sec:besfs-impl}

\begin{figure}[t]
\begin{center}
\includegraphics[width=8cm]{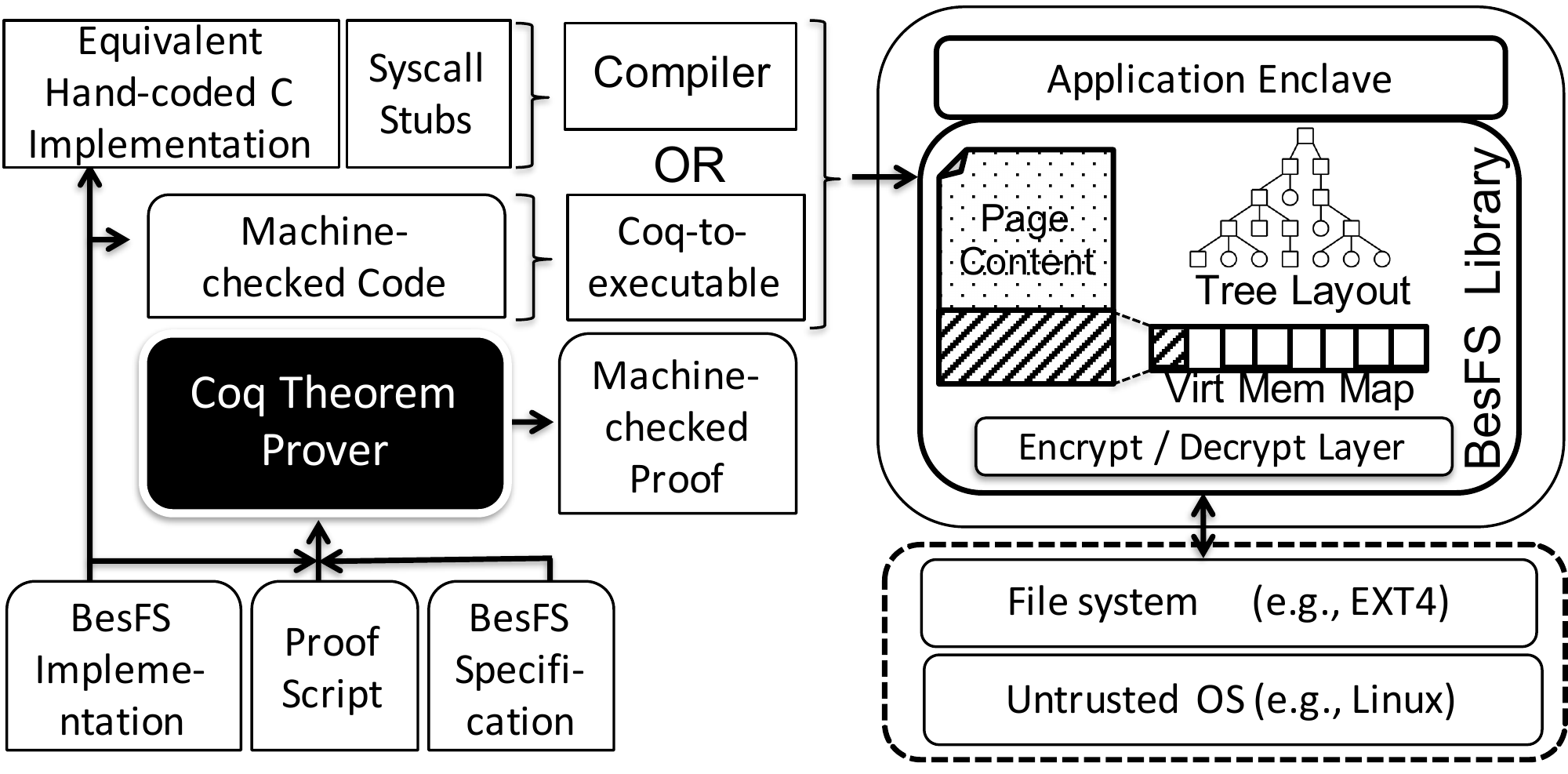}
\end{center}
\vspace{-15pt}
\caption{\codename Overview. Thick and dotted represents trusted and
untrusted components respectively.}
\vspace{-15pt}
\label{fig:overview}
\end{figure}

\codename defines a collection of data structures that implement the 
\codename interface design in Section~\ref{sec:interface}. Our
implementation in \coq is mechanically proof-checked and is the first
such system of its kind for enclaves. We build \codename types by
composition and/or induction over pre-defined \coq types $\tt{ascii}$,
$\tt{list}$, $\tt{nat}$, $\tt{bool}$, $\tt{set}$, $\tt{record}$,
$\tt{string}$, $\tt{map}$ in \coq libraries. All files and directories
in \codename have ids \fidi and \didi respectively. These ids are
mapped to the corresponding file and directory nodes \fdatai and
\ddatai. Specifically, \fdatai stores the file name, permissions, all
the pages that belong to this file, and the size of the file; \ddatai
stores the directory name, permission bits, and the number of files
and directories inside it. \metai represents the permissions and size
metadata. We give their simplified definitions:
\begin{align*}
\fidi &\coloneqq \mathbb{N} & \didi &\coloneqq \mathbb{N}\\
\pagei &\coloneqq [\mathtt{Byte}]_{\textsc{Pg\_Size}} & \permissioni &\coloneqq \text{W} \cross \text{R} \cross \text{E}\\
\metai &\coloneqq \permissioni \cross \mathbb{N} & \pageidi &\coloneqq \mathbb{N}\\
\fdatai &\coloneqq \mathtt{Str} \cross \metai \cross [\pageidi] & \ddatai &\coloneqq \mathtt{Str} \cross \metai\\
\treei &\coloneqq \textsc{File:}\; \fidi \mid \textsc{Dir:}\; \didi \cross [\treei] & \ohi & \coloneqq [\fidi \cross \mathbb{N}] \; \mhi \coloneqq [\mathbb{N} \cross \mathbb{N}] 
\end{align*}
The \codename filesystem layout \treei stores \fidi and \didi in a
tree form to represent the directory tree structure. The list of open
file handles \ohi stores tuples of \fidi and cursor position. Lastly,
each page is a sequence of \textsc{Pg\_Size} bytes which is the
typical size of a page\footnote{We set the page size
(\textsc{Pg\_Size}) to $4096$ bytes.} and has a unique page number
\pageidi. Finally, the entire filesystem memory map is stored as a
list \vmi. \codename uses \vmi to track the metadata for each page
allocated outside the enclave to the filesystem. \vmi does not save
the actual page content of the file inside the enclave, but only saves
the metadata such as file id, page id, and AES-GCM authentication tags
(Figure~\ref{fig:overview}).
To summarize, \codename implementation state comprises of:
\begin{equation*}
\footnotesize
\filesys \coloneqq (\layouti: \treei, \fhi:\ohi, \mmhi:\mhi, \vmi:
[\pagei], \mmi: [\pagei], \fmapi: \fidi \rightarrow \fdatai, \dmapi:
\didi \rightarrow \ddatai) 
\end{equation*}
\codename implementation must satisfy the state properties
\SP{1}-\SP{5} and transition properties \TP{1}-\TP{15} outlined in
Section~\ref{sec:interface}. Table~\ref{tab:ds-form} summarizes the
enforced invariants. Next, we discuss how we achieve this for each
data structure.

\paragraph{Virtual Memory Map (\memory).} 
Each file is an ordered sequence of pages. \codename assigns page ids
to each page in the filesystem. \codename virtual memory map \memory
is completely independent and unrelated to the OS-allocated virtual
address. For \codename, the filesystem memory is represented by a set
of virtual memory pages. Each page is a sequence of \textsc{Pg\_Size}
bytes and is represented by a unique page id \pageidi. \memory tracks
the virtual memory layout by storing the page metadata in the
filesystem. $4000$ bytes of each page comprises of the page content
while the remaining $96$ bytes are metadata for integrity protection
and can be used to store other metadata currently not traced by
\besfs. Pages are stored outside the enclave in an encrypted form and
are decrypted at the enclave boundary. \codename uses the virtual
memory map \memory inside the enclave to track and verify the
integrity of the page content returned by the OS. This mechanism is
similar to merkle tree implementations for encrypted
filesystems~\cite{leveefs}. \codename further ensures that a page 
belongs only to a single file and files do not have page overlaps. The
\memory map implementation marks the unallocated page metadata slots
as free in the pool.

\paragraph{Anonymous Memory Mapping (\ammap) \& Handles (\mh).}
When an anonymously mmaped memory region is first allocated in the
untrusted memory, \codename first checks if the allocation is valid
i.e., the memory returned by the OS is indeed zeroed out. \codename
then makes a copy of it into its enclave protected memory. \footnote{
The scalability of such a virtual address space mapping duplication is
not affected by the current limit on the EPC size ($90$ MB), because
SGX does not limit the enclave virtual memory to $90$ MB.} During this
step, \codename registers a handle for the new mapping which consists
of the start address and the total length of the mapped memory.
\codename allocation ensures that the mmaped regions do not overlap
with existing allocations. All accesses to the mmaped region are
redirected to the protected memory. When the region is unmmaped,
\codename deletes the handle, marks the pages in protected memory as
available, and relays the unmap call to the OS. Further, it ensures
that the memory layout does not overlap after unmap. 

\paragraph{Files \& Directories (\nodemap).}
Each file's information including the file name, the current size, and
 the permission bits are stored in a file node \fdatai. Each file's
content is a sequence of bytes, partitioned into uniformly sized
pages. This content is tracked by keeping an ordered list of virtual
memory page ids [\pageidi]. For example, the first id in a file node's
page list points to the exact page in the virtual memory where the
first $n$ bytes of the page are stored. \codename maintains a map
\fmapi which associates each file node \fdatai with a unique file
identifier \fidi. Similar to file nodes, \codename has directory nodes
\ddatai to track directory information such as names and permissions.
Each directory is associated with a unique directory id \didi. The
directory map \dmapi tracks the relationship between ids and nodes.

\paragraph{Layout \& Paths (\pathset).} 
\codename tracks the paths for all files and directories via a tree
layout \treei. Each node in the tree can be a file node id \fidi or a
directory node id \didi. Files are leaf nodes and each directory can
have its own tree layout. \codename does not allow cycles in the tree
layout and all levels have non-duplicate directory/file names. 

\input{tables/ds-form-table}

\paragraph{Open File Handles (\oh).} 
Each open file has a file handle which is allocated when the file is
first opened. The file handle comprises the file id \fidi and the
current cursor position for that file. \codename tracks all the list
of open files via the open file handles list \ohi. All operations on
an open file are done via its file handle. When the file is closed,
the file handle is removed from the list. Further, the \ohi list
cannot have any duplicate \fidi because each open file can have only
one handle.

\paragraph{Good State.}
\codename must satisfy all the data structure invariants in
Table~\ref{tab:ds-form} before and after any interface invocation to
be in a good state. A state is good if the following holds true:
\begin{gather*}
\mathrm{NoDupName}(\layouti) \wedge 
\mathrm{NoDup}(\textsc{Fids}(\layouti)) \wedge 
\mathrm{NoDup}(\textsc{Dids}(\layouti)) \wedge \null \\
\mathrm{NoDup}(\textsc{Ids}(\fhi)) 
\wedge \exists\,d\,s\text{~s.t.~} \layouti = \textsc{Dir:}\; d\; s \wedge \null\\
\forall i\,j,\, i \neq j \Rightarrow \fmapi(i)[2] \cap \fmapi(j)[2] = \emptyset
\end{gather*}

\paragraph{Known Limitations.}
\codename implementation does not support a small set of filesystem
operations, such as symbolic links, which are unsafe as per our safety
properties. Although our currently \codename does not reason about
other metadata information such as time-stamps (e.g., $\tt{mtime}$,
$\tt{atime}$, $\tt{ctime}$). There is no fundamental limitation in
adding them to \codename for detecting potential attacks from a
malicious OS. SGX does not support shared memory between enclaves.
Typical enclave applications do not concurrently access protected
files. Thus, we do not consider multi-enclave or concurrent access to
shared enclave files. \codename enforces an atomicity property and
does not reason about APIs for explicit synchronization (e.g.,
$\tt{sync}$, $\tt{fsync}$, and $\tt{fdatasync}$).\footnote{For
non-explicit synchronization, the enclave has to explicitly invoke
them to ask the OS to persist the changes.} Nonetheless, it is
compatible with them and detects any violation by the OS. We have
consciously decided to not support these functionalities in our first
version of \codename to maintain simplicity.

%% file: tables/ds-form-table.tex
\begin{table}
\centering
\resizebox{0.48\textwidth}{!}{%
\begin{tabular}{lll}
\toprule
\begin{tabular}[c]{@{}l@{}}Virtual\\ Memory Map\end{tabular} 
 & \memory & $\forall i\,j,\, i \neq j \Rightarrow \fmapi(i)[2] \cap \fmapi(j)[2] = \emptyset$ \\ \midrule
\multirow{4}{*}{\begin{tabular}[c]{@{}l@{}}Files \&\\ Directories\end{tabular}} & \multirow{4}{*}{\nodemap} & $\textsc{Fids}(\textsc{File:}\; i)\coloneqq [i]$ \\
 & & $\textsc{Fids}(\textsc{Dir:}\; i\; s)\coloneqq \textsc{Fids}(s[1])+\dots+\textsc{Fids}(s[n])$ \\
 & & $\textsc{Dids}(\textsc{File:}\; i)\coloneqq []$ \\
 & & $\textsc{Dids}(\textsc{Dir:}\; i\; s)\coloneqq [i]+\textsc{Dids}(s[1])+\dots+\textsc{Dids}(s[n])$ \\ \midrule
\multirow{6}{*}{\begin{tabular}[c]{@{}l@{}}Layout \&\\ Paths\end{tabular}} & \multirow{6}{*}{\pathset} & $\textsc{TreeName}(\textsc{File:}\; i)\coloneqq \fmapi(i)[0]$ \\
 & & $\textsc{TreeName}(\textsc{Dir:}\; i\; s)\coloneqq \dmapi(i)[0]$ \\
 & & $\mathrm{NoDupName}(t:\treei) \coloneqq \exists\, i, t = \textsc{File:}\; i \vee \null$ \\
 & & $\exists\,d\,s, t = \textsc{Dir:}\; d \; s \wedge (\forall i, \, \mathrm{NoDupName}(s[i])) \wedge$ \\
 & & $(\forall i j,\, i \neq j \Rightarrow \textsc{TreeName}(s[i])\neq\textsc{TreeName}(s[j]))$ \\
 & & $\mathrm{NoDup}([\dots s_i\dots s_j \dots])\coloneqq \forall i\,j,\, i \neq j \Rightarrow s_i \neq s_j$ \\ \midrule
\multirow{2}{*}{\begin{tabular}[c]{@{}l@{}}Open file\\ handles\end{tabular}} & \multirow{2}{*}{\oh} & $ \textsc{Ids}([\dots,(f_i, p_i),\dots,(f_j,p_j),\dots]: \ohi)\coloneqq [\dots,f_i,\dots,f_j,\dots]$ \\
 & & $\mathrm{NoDup}(\textsc{Ids}[\dots s_i\dots s_j \dots])\coloneqq \forall i\,j,\, i \neq j \Rightarrow s_i \neq s_j$ \\ \midrule 
\multirow{2}{*}{\begin{tabular}[c]{@{}l@{}}Anon Mmaps\\ \& Handles\end{tabular}} & 
\multirow{2}{*}{\begin{tabular}[c]{@{}l@{}}$\ammapm$\\ $\mhm$\end{tabular}} &
\multirow{2}{*}{\begin{tabular}[c]{@{}l@{}} $\textsc{Mids}(\textsc{Q:}\; i)\coloneqq [i] \, \land \, \mathrm{NoDup}(\textsc{MIds}[\dots (a_i, l_i) \dots (a_j, l_j) \dots])\coloneqq $ \\ $\forall i\,j,\, ,i \neq j, \, \exists k \in (0, l_i) \Rightarrow a_j \neq a_i + k$ \end{tabular}} \\ \\ \bottomrule
\end{tabular}%
}

\caption{\codename data structures definitions \& invariants.}
\vspace{-15pt}
\label{tab:ds-form}
\end{table}

%% file: chapters/safety-proof.tex
\section{\codename Safety Proof \& Modeling Challenges}
\label{sec:proof}

The key theorems for our \codename implementation are that the
functions meet our interface specifications. For each method of our
interface, we must prove that the implementation satisfies the state
properties (\SP{1}-\SP{5}) from Section~\ref{sec:interface} and the
transition properties (\TP{1}-\TP{15}) outlined in
Table~\ref{tab:api}. We assume \codename is running on a hostile OS
that can take any actions permitted by the hardware.

As one can readily see, our implementation uses recursive data
structures and its state properties require second-order logic. For
example, the \codename filesystem layout \treei in
Section~\ref{sec:besfs-impl} is defined mutually recursively in
terms of a forest (a list of trees). This motivates our choice of
\coq, an interactive proof assistant supporting calculus of inductive
constructions. \coq allows the prover to write definitions of data
structures and interface specification in a language called \gallina,
which is a purely functional language. The statements of the theorems
are written in \gallina as well. The proofs of the statement, called
{\em proof scripts} are written in a language called \ltac. \ltac's
library of tactics, or one-line commands, encode standard proof
strategies for ease of writing proofs. 

\paragraph{Purely Functional.}
The programming language provided by \coq is purely functional, having
no global state variables. However, the filesystem is inherently
stateful. So, we use {\em state passing} to bridge this gap. The state
resulting from the operation of each method is explicitly passed as a
parameter to the next call. If we explicitly pass these state in each
call, it is prone to clutter and accidental omission; therefore, we
define them as a monad. As we can see in the definition of
$\tt{fs\_write}$, the code is purely functional but it looks like
the traditional imperative program. The benefit of this monadic style
programming is that it hides the explicit state passing, which makes
the code more elegant and less error-prone.

While proof script checking, if \coq encounters a memoized expression
for $f(z)$, it will skip proving $f(z)$ again. This is a challenge
because in a sequence of system calls the same call to $f$ with
identical arguments may return different values. Therefore, we have to
force \coq to treat each call as different. To implement this, we
introduce an implicit counter as an argument to all the calls. It 
increments after each call completes. For example, consider the
consecutive external calls $\tt{read\_dir}$, $\tt{create\_dir}$, and
$\tt{read\_dir}$. The two $\tt{read\_dir}$ commands may read the same
directory (the same argument) but with different return values because
of the $\tt{create\_dir}$ command. To reason about such cases, the
real arguments passed to the external calls contain not only the
common arguments but also an ever-increasing global counter. Thus, in
our $\tt{read\_dir}$ example, the two commands with original argument
$p$ will be represented as $\tt{read\_dir}(p, n)$ and
$\tt{read\_dir}(p, n + 1)$ so that \coq treats them as different.

\paragraph{Atomicity}.
The purely functional nature of \coq proofs helps to prove the
atomicity of each method call. In an enclave, its internal state is
not accessible by the OS even if it gets interrupted; so, in a way,
the enclave behaves like a pure function between two OS calls. This
simplifies our proof for atomicity. We structure the proof script to
check if an error state is reachable from the input state and the
OS-returned values; if so, the input state is retained as the output
state. If no error is possible, the output state is set to the new
state. As a concrete example, the $\tt{write}$ method progressively
checks $5$ conditions (1: argument $\tt{id}$ is in the handler; 2: the
specified position is correct; 3: iut writes to the copied virtual
memory successfully; 4: the external call to $\tt{seek}$ succeeds; and
5: the external call to $\tt{write}$ succeeds.) before changing the
state.

\paragraph{Non-deterministic Recursive Termination.}
\gallina guarantees that any theorem about a \gallina program is
consistent, i.e., it cannot be both proved and disproved. Further, all
programs in \gallina must terminate, since the type of the program is
the statement of a theorem.\footnote{A non-terminating program such as
$\mathtt{let} f(x) := f(x)$ has an arbitrary type, and hence any
theorem is valid about it.} \coq uses a small set of syntactic
criteria to ensure the termination. \gallina's termination requirement
poses challenges for writing a \codename implementation, which uses
recursive data structures. In most cases, the termination proof for
\codename properties are automatic; however, for a small number of
properties, we have to provide an explicit termination proof. For
instance, $\tt{write\_to\_buffer}$ does not admit a syntactic
check for termination, as there is a recursive call. To prove
termination, via induction, we show that the input buffer size 
strictly reduces for each invocation of $\tt{write}$. Effectively, we
establish that there are no infinite chains of nested recursive calls.

\paragraph{Mutually Recursive Data Structures.}
Most of our data structure proofs are by induction and \coq always
provides an induction scheme for each inductively declared structure.
The automatically generated induction scheme from \coq is {\em not}
always strong enough to prove some of our properties. Specifically, a
key data structure in our design is a tree, the leaves of which are a
list of trees---this represents the directory and file layouts
(Section~\ref{sec:interface})---in this case. 
\input{listings/mutually-recursive-proof}
We provide an inductive statement $\tt{Tree\_ind2}$ that is stronger
than \coq-provided induction scheme $\tt{Tree\_ind}$, shown in the
above listing. $\tt{Tree\_ind}$ is correct but useless. We dispatch
the proof by the principle of strong induction, which is
$\tt{Tree\_ind2}$. Our induction property uses \coq's second-order
logic capability, as the above code listing shows that the
sub-property $\tt{P}$ is an input argument to the main property. In
our full proof, a number of specific properties instantiate $\tt{P}$.

\paragraph{External Calls to the OS.}
We assume that calls to the OS always terminate to allow
\coq to provide a proof. If the call terminates, the safety is
guaranteed; the OS can decide not to terminate which constitutes as
denial-of-service.

\paragraph{Odds \& Ends.}
Out of the $167$ lemmas, we prove $75$ of them using inductions and
the rest of them by logical deductions. There are two kinds of
inductions in our proofs: strong induction and weak induction, the
difference is the proof obligation. For example, in weak induction we
need to prove: if $P(k)$ is $\true$ then $P(k + 1)$ is $\true$. In
strong induction, it is: if $P(i)$ is $\true$ for all $i$ less than or
equal to $k$ then $P(k+1)$ is $\true$. Our customized induction
principle for $\tt{Tree}$ is a typical strong induction. In all, we
proved $75$ lemmas by induction ($39$ and $36$ lemmas by strong  and
weak induction respectively).

We do not implement $\tt{get\_next\_free\_page}$ but enforce that an
implementation must satisfy the property that the new page allocated
by the function is not used for existing files and is a valid page
(less than the upper bound limit). Similarly, for functions
$\tt{new\_fid}$ and $\tt{new\_did}$ we enforce the new ids are
unique to avoid conflict. It is formally stated as
$\mathtt{new\_fid}(\layouti) \not\in\textsc{Fids}(\layouti)$ and
$\mathtt{new\_did}(\layouti) \not\in\textsc{Dids}(\layouti)$
respectively. Note that we only give a specification for allocating
new pages and ids for files and directories because we do not want to
restrict the page and namespace management algorithm. This way, the
implementation can use a naive strategy of just allocating a new
id/page for each request, employ a sophisticated re-use strategy to
allocated previously freed ids, or use temporal and spatial
optimizations for page allocation as long as they fulfill our safety
conditions.

%% file: listings/mutually-recursive-proof.tex
\begin{lstlisting}[language={}, xleftmargin=0.3cm]
Tree_ind: forall P: Tree -> Prop,
 (forall f: Fid, P (Fnode f)) -> (forall (d: Did) (l: list Tree), 
 P (Dnode d l)) -> forall t: Tree, P t
Tree_ind2: forall P : Tree -> Prop,
 (forall f: F, P (Fnode f)) -> (forall (d: Did) (l: list Tree), 
 forall P l -> P (Dnode d l)) -> forall t: Tree, P t
\end{lstlisting}

%% file: chapters/coq-to-exec.tex
\section{\coq to Executable Code}
\label{sec:extraction}

\codename ~\coq definitions and proof script comprise $4625$ LOC with
$167$ lemmas and $2$ main theorems. The development effort for
\codename was approximately two-human years for designing the
specifications and proving them. The \coq implementation has a
machine-checked proof of correctness, i.e., matching the
specification. The \coq code, however, needs to be converted to
executable code to run in an enclave. Currently, \coq supports
automatic extraction to three high-level languages: OCaml, \haskell,
and Scheme~\cite{coq-extraction}. We can successfully compile our code
to executables; however, none of these three functional languages have
runtime support for Intel SGX, primarily due to the lack of a memory
manager (e.g., garbage collector) that is compatible with SGX.

Further, we have tried to run our compiled code in these three
languages on existing library OSes with SGX, but without success. 
Specifically, we find that two state-of-the-art frameworks,
Graphene-SGX~\cite{graphene-sgx} and \panoply~\cite{panoply}, are not
robust enough to run compiled \haskell or OCaml ``hello world''
programs. Our investigation reveals that supporting these functional
language runtimes in entirety would require extensive foundational
work, such as porting memory managers, and SGX support on existing
library OSes misses several critical OS abstractions. Specifically,
Graphene-SGX does not support $\tt{create\_timer}$, $\tt{set\_timer}$,
$\tt{delete\_timer}$, and $\tt{sigaction}$ syscalls. We attempted to
add support for these syscalls, but it is a non-trivial amount of work
to add support for an entire subsystem to Graphene-SGX. In
Section~\ref{sec:discussion}, we discuss why certified compilation
from \coq to machine code is currently not practical, but a promising
future direction.

With no publicly available enclave system supporting compiled programs
for high-level language that \coq extracts to, we resorted to a manual
line-by-line translation of our machine-checked \coq implementation to
$\tt{C}$ code. Our $\tt{C}$ implementation comprises of $863$ LOC core
logic and $586$ LOC helper functions, totaling $1449$ LOC
(Table~\ref{tab:component-loc}). Our \coq code intentionally leaves
out the implementation of untrusted POSIX calls. At enclave runtime,
these calls have to be redirected to an actual filesystem provided by
the OS (whose behavior is not trusted).

\input{tables/component-loc}

\paragraph{Ease of Integration}. 
Our $\tt{C}$ implementation can be integrated with any SGX
framework~\cite{panoply, graphene-sgx, scone} as well as stand-alone
SGX applications~\cite{privado} and SGX SDK~\cite{sgx-sdk} (See
Section~\ref{sec:case-studies}). We choose \panoply as the SGX 
framework to integrate and test \besfs. For adding \besfs support, we
wrap the application's file system calls and marshal its arguments to
make them compatible with \besfs interface described in
Section~\ref{sec:interface}. Once \panoply collects the return values
from the external \libc call, we unmarshal the return values and give
it back to \codename. \codename checks the return values and our
wrapper then converts back the results to a data type expected by the
application. If \besfs deems the results as safe we return the final
output of the API call to the application, else we flag a safety
violation. We add $724$ LOC to \panoply. 

%% file: tables/component-loc.tex
\begin{table}[]
\centering
\resizebox{0.38\textwidth}{!}{%
\begin{tabular}{@{}lcrl@{}}%
\toprule
\multicolumn{1}{c}{\textbf{Component}} & \multicolumn{1}{c}{\textbf{Language}} & \textbf{LOC} & \textbf{Size (in KB)} \\ \midrule
\multicolumn{4}{c}{\textbf {\em Machine-proved Implementation}} \\ 
\coq definitions \& Proofs & \gallina & 3676 & 1757.38 \\ 
\multicolumn{4}{c}{\textbf {\em Hand-coded Implementation}} \\ 
Implementation & C & 863 & 172.39 \\ 
External Call Interface & C & 469 & 201.55 \\ 
SGX Utils & C & 117 & 667.04 \\ \midrule
\multicolumn{2}{l}{\textbf{Total}} & \textbf{1449} & \textbf{1040.98} \\ \bottomrule
\end{tabular}
}
\caption{LOC for various components of \codename.}
\vspace{-15pt}
\label{tab:component-loc}
\end{table}

%% file: chapters/eval.tex
\section{Evaluation}
\label{sec:eval}

Our evaluation goal is to demonstrate the following:

\begin{itemize}
\squish
\item
\codename safety definition is compatible with the semantics of POSIX
APIs expected by benign applications.
\item 
Our API has the right abstraction and is expressive enough to support
a wide range of applications.
\item 
The bugs uncovered in our implementation due to \codename formal
verification efforts.
\item 
\codename can be integrated into a real system.
\item 
Performance of \codename for (a) I/O intensive benchmarks; (b) CPU
intensive benchmarks; (c) per-call latencies for files; and (d)
real-world application workloads in typical enclave deployments.
\end{itemize}

\paragraph{Experimental Setup.}
All our experiments were conducted on a machine with Intel Skylake
i7-6600U CPU ($2.60$ GHz, $4$ cores) with $12$ GB memory and $128$ MB
EPC of which $96$MB is available to user enclaves. We execute our
benchmark on Ubuntu $18.04$ LTS. We use our hand-coded $\tt{C}$
implementation of \codename and \panoply (unless stated otherwise) to
run our benchmarks in an enclave. \panoply internally uses Intel SGX
SDK Linux Open Source version $2.4$~\cite{sgx-sdk}.~\footnote{We have
also benchmarked \codename on Ubuntu $14.04$, SGX SDK $1.6$.}
\codename uses ext4 as the underlying POSIX compliant filesystem.

\paragraph{Benchmarks Selection Criteria \& Description.}
Our selection is aimed at showcasing how well \codename fares in
reaching its design goals. Since our evaluation goals for \codename
are multi-faceted, we selected a wide variety of micro-benchmarks,
benchmarks, and real-world applications.
First, we use the micro-benchmark suite from \fscq~\cite{fscq}. It
comprises workloads to test each file-related system call via 
different sequences of filesystem operations on large and small files.
Second, we use IOZone~\cite{iozone}, a well-known and a broad
filesystem benchmark for measuring bandwidth for different
file access patterns with $13$ tests for $7$ standard
operations. Third, for testing \codename on non-I/O intensive
applications, we use CPU-intensive programs from SPEC
CINT2006~\cite{spec}. We were able to port $7/12$ programs from SPEC.
We were unable to port the rest of the benchmarks because some
programs from SPEC ($\tt{omnetpp}$, $\tt{perlbench}$,
$\tt{xalancbmk}$) use non-$\tt{C}$ APIs which are not supported in
\panoply. Other limitations such as lack of support for $\tt{longjmp}$
in \panoply's SDK version prevent us from running the $\tt{gobmk}$ and
$\tt{gcc}$ programs. Fourth, we use all applications from \panoply---a
system to execute legacy applications in enclaves. These $4$
real-world applications (H2O web server, TDS database client, OpenSSL
library, and Tor) have a mix of CPU, memory, and file, and network IO
workloads. We successfully port $3/4$ case-studies to \codename (see
Section~\ref{sec:case-studies} for details) and use the same workloads
as that in \panoply~\cite{panoply}. Lastly, we select all the $10$
real-world applications from \privado~\cite{privado} which perform
inference over CIFAR10 and ImageNet using state-of-the-art neural
network models. 
Thus, our final evaluation is on a total of $31$ applications: (a)
$10$ programs from \fscq for micro-benchmarking per-call latencies for
file operations, (b) IOZone and $7$ programs from SPEC for measuring
the overhead of \codename on IO-intensive and CPU-intensive
benchmarks. (c) $3$ applications from \panoply and $10$ applications
from \privado for demonstrating the effect of \codename on real-world
enclave usage. All our results are aggregated over $5$ runs. 

\input{chapters/compatibility}
\input{chapters/bugs}
\input{chapters/perf}
\input{chapters/case-studies}

%% file: chapters/compatibility.tex
\subsection{Expressiveness \& Compatibility}
\label{subsec:eval-compat}

\input{tables/ltrace-stats-new}

We empirically demonstrate that if the underlying filesystem and the OS
are POSIX compliant and benign then \codename is not overly
restrictive in the safety conditions. We first analyze all syscalls
and \libc calls made by our benchmarks for various workloads using
$\tt{strace}$ and $\tt{ltrace}$ respectively. We then filter out the
fraction of filesystem related calls. Table~\ref{bm-stats} shows
the statistics of the type of filesystem call and its frequency for
our benchmarks. We observe a total of $235008$ filesystem calls
comprising of $18$ unique APIs. \codename can protect $235000/235008$
of them. \input{tables/api-composition}

\paragraph{Compositional Power of BesFS.}
\codename directly reasons about $15$ calls using the core APIs
outlined in Section~\ref{sec:interface}. We use \codename's
composition theorem and support additional $22$ auxiliary APIs that
have to be intercepted such that \codename checks all the file
operations for safety. For example, $\tt{fgets}$ reads a file and
stops after an EOF or a newline. The read is limited to at most one
less character than $\tt{size}$ parameter specified in the call. We
implement $\tt{fgets}$ by using \codename's core API for read (see 
Table~\ref{api-composition}). Since we do not know the location of the
newline character, we read the input file character-by-character and
stop when we see a new line, EOF, or if the buffer size reaches the
value $\tt{size}$. Similarly, we already know the total size of the
buffer when writing the content to the output file (e.g., after
resolving the format specifiers in $\tt{fprintf}$). Thus we write the
complete buffer in one single call. \libc calls use flags to dictate
what operations the API must perform. For example, the application can
use the $\tt{fopen}$ API to open a file for writing. If the
application specifies the append flag ($\tt{``a"}$), the library
creates the file if it does not exist and positions the cursor at the
end of the file. To achieve the same functionality using \codename, we
first try to open the file, if it fails with an $\tt{ENOENT}$ error,
we check if the parent directory exists. If so, we first create a new
file. If the file exists, we open the file and then explicitly seek
the cursor to the end of the file. We implement and support a total of
$16$ flags in total for our $3$ APIs which require flags. Our
implementation currently supports the common flags used by
applications and can be extended in the future using our core APIs.

\codename does not reason about the safety of the remaining $2$ APIs
which amount to a total of $8$ calls in our benchmarks. Although
\codename does not support these unsafe calls, it still allows the
enclave to perform those calls. Importantly, these unsupported calls
do not interfere with the runs in our test suite and do not affect our
test executions. By the virtue of \codename's atomicity property,
synchronization calls such as $\tt{sync}$, $\tt{fsync}$, and
$\tt{fdatasync}$ have to be implicitly invoked for the OS after each
function call to persist the changes. We experimentally confirm that
the program produces the same output with and without \codename, thus
reaffirming that our safety checks do not alter the program behavior.

%% file: tables/ltrace-stats-new.tex
\begin{table}[]
\centering
\resizebox{0.48\textwidth}{!}{%
\begin{tabular}{@{}lrrrrrrrrrr@{}}%
\toprule
\multicolumn{1}{c}{\multirow{2}{*}{\textbf{\begin{tabular}[c]{@{}c@{}}LibC\\ Calls\end{tabular}}}} & \multicolumn{7}{c}{\textbf{SPEC CINT 2006}} & \multicolumn{2}{c}{\textbf{\fscq}} & \multirow{2}{*}{\textbf{Total}} \\ \cmidrule(l){2-8} \cmidrule(l){9-10}
\multicolumn{1}{c}{} & \multicolumn{1}{c}{\textbf{astar}} & \multicolumn{1}{c}{\textbf{mcf}} & \multicolumn{1}{c}{\textbf{bzip2}} & \multicolumn{1}{c}{\textbf{hmmer}} & \multicolumn{1}{c}{\textbf{libqu}} & \multicolumn{1}{c}{\textbf{h264}} & \multicolumn{1}{c}{\textbf{sjeng}} & \multicolumn{1}{c}{\textbf{small}} & \multicolumn{1}{c}{\textbf{large}} & \\ \midrule
\multicolumn{11}{c}{{\codename Core Calls}} \\ 
\midrule
{open} & 3 & 0 & 1 & 0 & 0 & 7 & 0 & 2 & 1 & 14 \\ 
{read} & 27 & 0 & 4 & 0 & 0 & 129 & 0 & 1 & 3072 & 3233 \\ 
{write} & 0 & 0 & 0 & 0 & 0 & 0 & 0 & 1 & 66560 & 66561 \\ 
{lseek} & 0 & 0 & 0 & 0 & 0 & 75 & 0 & 0 & 66563 & 66638 \\ 
{remove} & 0 & 0 & 0 & 0 & 0 & 0 & 0 & 2 & 1 & 3 \\ 
{close} & 3 & 0 & 1 & 0 & 0 & 7 & 0 & 2 & 1 & 14 \\ 
{mkdir} & 0 & 0 & 0 & 0 & 0 & 0 & 0 & 100 & 0 & 100 \\ \midrule
\multicolumn{11}{c}{{\codename Auxiliary Calls}} \\ \midrule
{fopen} & 1 & 2 & 0 & 5 & 0 & 6 & 1 & 0 & 0 & 15 \\ 
{fread} & 1 & 0 & 0 & 1 & 0 & 1 & 0 & 0 & 0 & 3 \\ 
{fwrite} & 0 & 1035 & 0 & 6 & 0 & 13 & 2 & 0 & 0 & 1056 \\ 
{fgets} & 0 & 90435 & 0 & 108 & 0 & 0 & 5 & 0 & 0 & 90548 \\ 
{fscanf} & 12 & 0 & 0 & 0 & 0 & 24 & 0 & 0 & 0 & 36 \\ 
{fprintf} & 0 & 5985 & 0 & 605 & 0 & 17 & 162 & 0 & 0 & 6769 \\ 
{fseek} & 0 & 0 & 0 & 0 & 0 & 2 & 0 & 0 & 0 & 2 \\ 
{ftell} & 0 & 0 & 0 & 4 & 0 & 1 & 0 & 0 & 0 & 5 \\ 
{rewind} & 0 & 0 & 0 & 3 & 0 & 0 & 0 & 0 & 0 & 3 \\ \midrule
\multicolumn{11}{c}{{Unsafe Calls}} \\ \midrule
{fsync} & 0 & 0 & 0 & 0 & 0 & 0 & 0 & 0 & 2 & 2 \\ 
{rename} & 0 & 0 & 0 & 0 & 0 & 0 & 6 & 0 & 0 & 6 \\ \midrule
\textbf{Total} & 47 & 97457 & 6 & 732 & 0 & 282 & 176 & 108 & 136200 & \textbf{235008} \\ \bottomrule
\end{tabular}
}
\caption{
Frequency of filesystem calls. Rows $3-11$ and $13-22$ represent the
frequency of core and auxiliary calls supported by \codename
respectively. Rows $24-26$ show the frequency of unsafe calls for
each of our benchmarks.
}
\vspace{-15pt}
\label{bm-stats}
\end{table}

%% file: tables/api-composition.tex
\begin{table}[]
\centering
\resizebox{0.45\textwidth}{!}{%
\begin{tabular}{@{}lrccccccccccccc@{}}%
\toprule
\multicolumn{1}{c}{\multirow{2}{*}{\textbf{\begin{tabular}[c]{@{}c@{}}Libc \\ API\end{tabular}}}} & \multirow{2}{*}{\textbf{LOC}} & \multicolumn{13}{c}{\textbf{\codename Core API used for composition of LibC API}} \\ \cmidrule(l){3-15} 
\multicolumn{1}{c}{} & & \rotatebox{90}{fstat} & \rotatebox{90}{read} & \rotatebox{90}{open} & \rotatebox{90}{close} & \rotatebox{90}{seek} & \rotatebox{90}{create} & \rotatebox{90}{mkdir} & \rotatebox{90}{rmdir} & \rotatebox{90}{remove} & \rotatebox{90}{chmod} & \rotatebox{90}{readdir} & \rotatebox{90}{truncate} & \rotatebox{90}{write} \\ \midrule
{read} & 7 & & \checkmark & & & & & & & & & & & \\ \hline
{fread} & 25 & & \checkmark & & & & & & & & & & & \\ \hline
{fscanf} & 34 & & \checkmark & & & & & & & & & & & \\ \hline
{fwrite} & 12 & \checkmark & & & & & & & & & & & & \checkmark \\ \hline
{write} & 20 & \checkmark & & & & & & & & & & & & \checkmark \\ \hline
{fprintf} & 15 & \checkmark & & & & & & & & & & & & \checkmark \\ \hline
{fopen} & 78 & \checkmark & & \checkmark & & \checkmark & \checkmark & & & & & & \checkmark & \\ \hline
{open} & 60 & \checkmark & & \checkmark & & \checkmark & \checkmark & & & & & & \checkmark & \\ \hline
{fclose} & 9 & & & & \checkmark & & & & & & & & & \\ \hline
{close} & 17 & & & & \checkmark & & & & & & & & & \\ \hline
{fseek} & 31 & \checkmark & & & & \checkmark & & & & & & & & \\ \hline
{lseek} & 39 & \checkmark & & & & \checkmark & & & & & & & & \\ \hline
{rewind} & 5 & & & & & \checkmark & & & & & & & & \\ \hline
{creat} & 30 & & & \checkmark & & & \checkmark & & & & & & & \\ \hline
{mkdir} & 25 & & & & & & & \checkmark & & & & & & \\ \hline
{unlink} & 21 & & & & & & & & & \checkmark & & & & \\ \hline
{chmod} & 23 & & & & & & & & & & \checkmark & & & \\ \hline
{ftruncate} & 5 & & & & & & & & & & & & \checkmark & \\ \hline
{ftell} & 12 & \checkmark & & & & & & & & & & & & \\ \hline
{fgetc} & 9 & & \checkmark & & & & & & & & & & & \\ \hline
{fgets} & 25 & & \checkmark & & & & & & & & & & & \\ \hline
{readdir} & 10 & & & & & & & & & & & \checkmark & & \\ \bottomrule
\end{tabular}%
}
\caption{
Expressiveness of \codename. Row represents a \libc API used by our
benchmarks. Column $2$ represents the LOC added to implement the \libc
API. Columns $3-15$ represent the $13$ core APIs supported by
\codename. \checkmark represents that the API is used to compose
\libc API.
}
\vspace{-15pt}
\label{api-composition}
\end{table}

%% file: chapters/bugs.tex
\subsection{Do Proofs Help in Eliminating Bugs?}
\label{subsec:eval-bugs}

We encountered many mistakes and eliminated them during the
development as a part of our proof experience. This highlights the
importance of a machine-proved specification.

\input{chapters/graphs-combined}

\paragraph{Example 1: $\tt{seek}$ Specification Bug.}
In at least two of our functions, we need to test whether the position
of the current cursor is within the range of the file, in other words,
less than the length of the file. If the cursor is beyond the scope of
a specific file, any further operation such as read or write is
illegal. In the early versions of our \coq implementation, we simply
put ``$\tt{if~ pos < size}$'' as a judgment. But during the proof, we
found we cannot prove certain assertions because we had ignored the
corner case by mistake: when the file is just created with $0$ bytes
size, the only valid position is also $0$. 

\paragraph{Example 2: $\tt{write}$ Implementation Bug.}
\codename's $\tt{write}$ function input includes the position
($\tt{pos}$) at which the buffer is to be written. In our initial
\coq implementation of $\tt{write}$, we used the name $\tt{pos}$
for the cursor stored in the open handles (\oh). Thus, we had two
different variables being referred to by the same name. As a result,
the second variable value (the cursor) shadowed the write position.
This bug in $\tt{write}$ was violating the specification for the
argument $\tt{pos}$. We uncovered it when our proof was not going
through. However, once we fixed the bug by renaming the input
argument, we were able to prove the safety of $\tt{write}$.

\paragraph{Example 3: \panoply \& Intel SGX SDK Overflow Bugs.}
\panoply's $\tt{fread}$ and $\tt{fwrite}$ calls pass the size of
the buffer and a pointer to the buffer. \codename piggybacks on these
\panoply calls to read and write encrypted pages. While integrating
\codename code in \panoply, our integrity checks after read / write 
calls were failing. On further inspection, we identified stack
corruption bugs in both $\tt{fread}$ and $\tt{fwrite}$
implementations of \panoply. Specifically, if the buffer size is
larger than the maximum allowed stack size in the enclave
configuration file ($> 64$ KB in our experiments), even if we
pass the right buffer size, the enclave's stack is corrupted. To fix
this issue, we changed the SDK code to splice the buffer into smaller
sizes ($< 64$ KB) to read / write large buffers. After our fix,
the implementation passed \codename checks.

\paragraph{Example 4: \panoply Error Code Bugs.}
According to $\tt{fopen}$ POSIX specification, the function fails with
$\tt{ENOENT}$ if the filename does not name an existing file or is an
empty string. When we used \panoply's $\tt{fopen}$ interface, it did
not return the expected error code when the file did not exist. Our
\codename check after the external call flagged a warning of a safety
condition violation because \codename did not have a record of this
file but the external call claimed that the file existed. On
investigation, we discovered that \panoply had a bug in its
$\tt{errno}$ passing logic. In fact, on further testing of other
functions using \codename, we found $7$ distinct functions where
\panoply's error codes were incorrect. We tested against the $7$
attacks / bugs in \panoply after integrating \codename to ensure that
it did not violate any invariants.

\paragraph{Simulating a Malicious OS.}
First, we hand-crafted a suite of around $687$ tests cases in the form
of $\tt{assert}$ statements embedded in $40$ test-driver $\tt{C}$
programs that make a series of filesystem calls. To generate these
asserts and test drivers, we took our proof invariants and
systematically generated asserts which checked the given constraint.
We then coded the tests along with the assert statements. Second, to 
simulate the malicious OS, we manually crafted and planted known-bad
return values at the system call interface. We semi-randomly generated
these values, similar to SibylFS~\cite{sibylfs}. When simulating the
OS, it does not matter if the victim binary is executing inside or
outside of an enclave. This observation simplified our testing setup. 
For a clean way to hook on the syscalls and libc calls made by our
victim test-driver programs, we used the $\tt{ld\_preload}$
environment variable to optionally link the test case victim binaries
with our malicious syscall and libc return values.\footnote{Another
way is to write a malicious Linux kernel module to intercept calls
made by the victim enclave binary.} We then performed three sets of
executions of the victim binaries: (a) without our malicious library
and without \codename for ensuring that the victim binary executes in
the baseline case and recording the benign path for a given input; (b)
with our malicious library but without \codename to show that the lack
of checks causes the victim binary to execute unintended paths i.e.,
assertion failures; (c) with our malicious library and \codename to
check if \codename can detect the bad return values. We investigated
the resulting assertion failures in these runs. We report that all of
the failures observed in (b) but not (a) were due to lack of checks;
while they did not occur in case (c). This shows that \codename
invariants were able to prune all the planted bad return values.

%% file: chapters/graphs-combined.tex
\begin{figure*}[h]
\centering
\begin{minipage}[h]{0.24\textwidth}
\epsfig{file=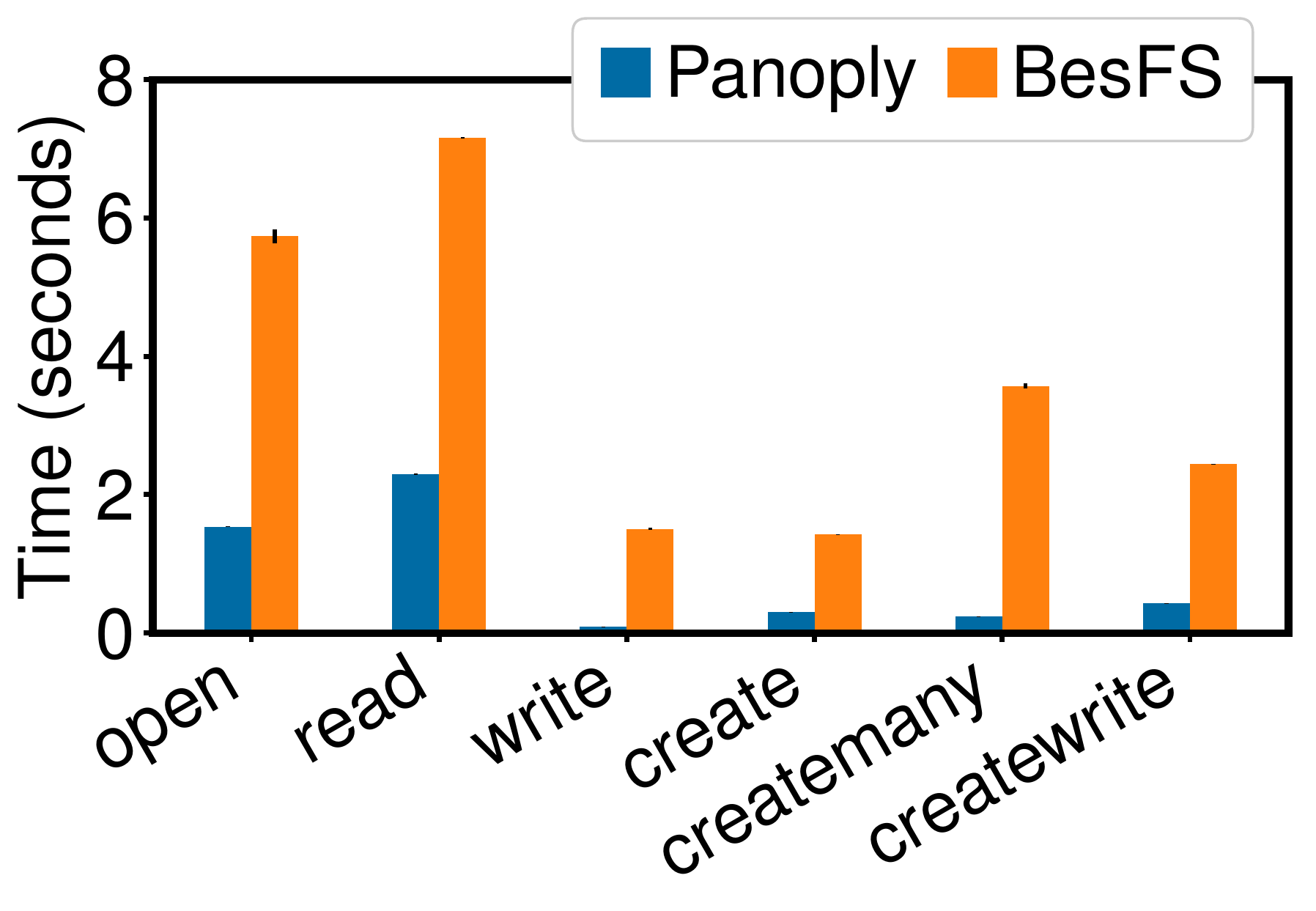, scale=0.22}
\subcaption{ \fscq Single Syscalls.}	
\label{fig:single}
\end{minipage}
\begin{minipage}[h]{0.24\textwidth}
\epsfig{file=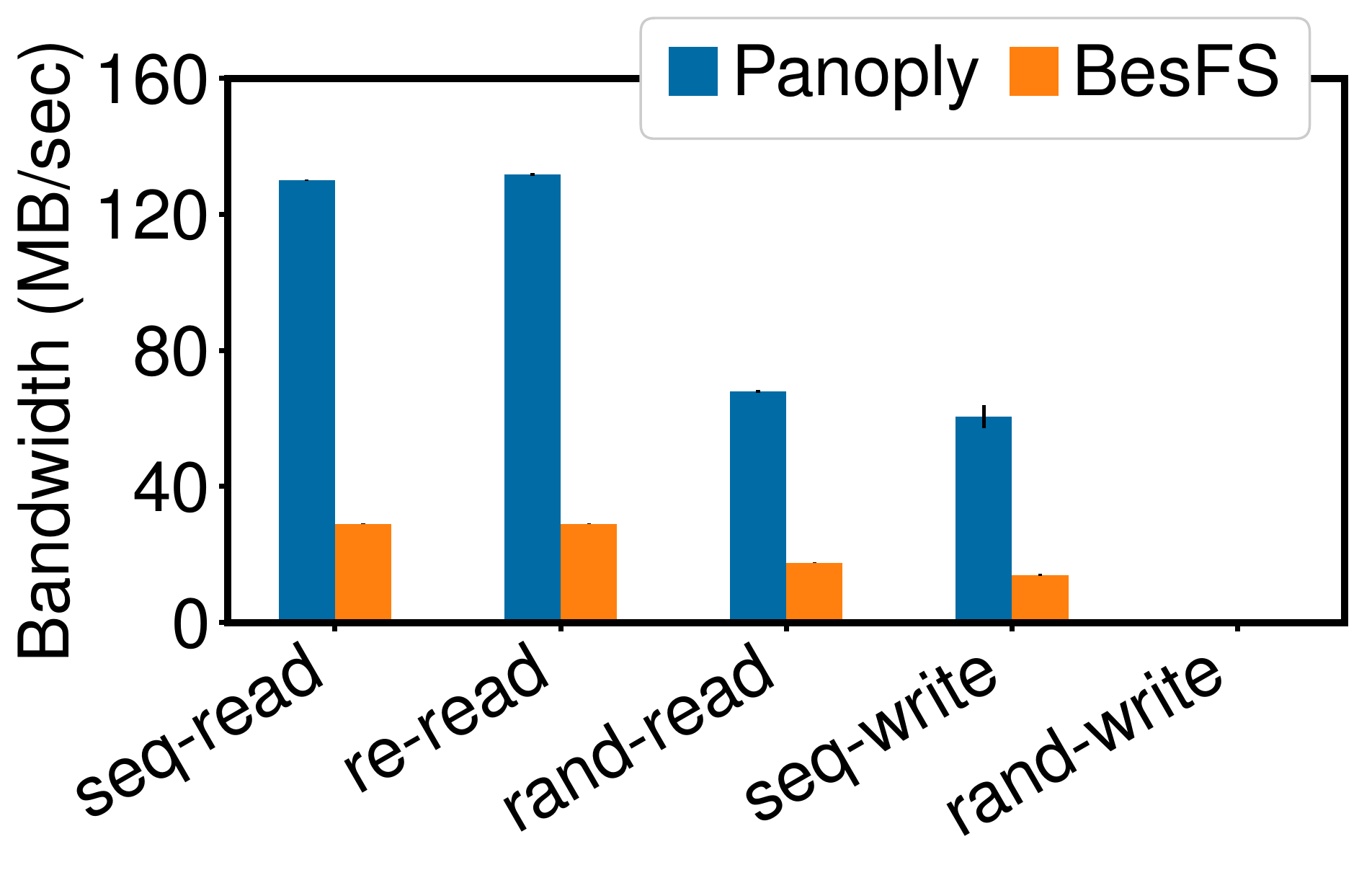, scale=0.22}
\subcaption{\fscq Large IO.}	
\label{fig:multi}
\end{minipage}
\begin{minipage}[h]{0.24\textwidth}
\epsfig{file=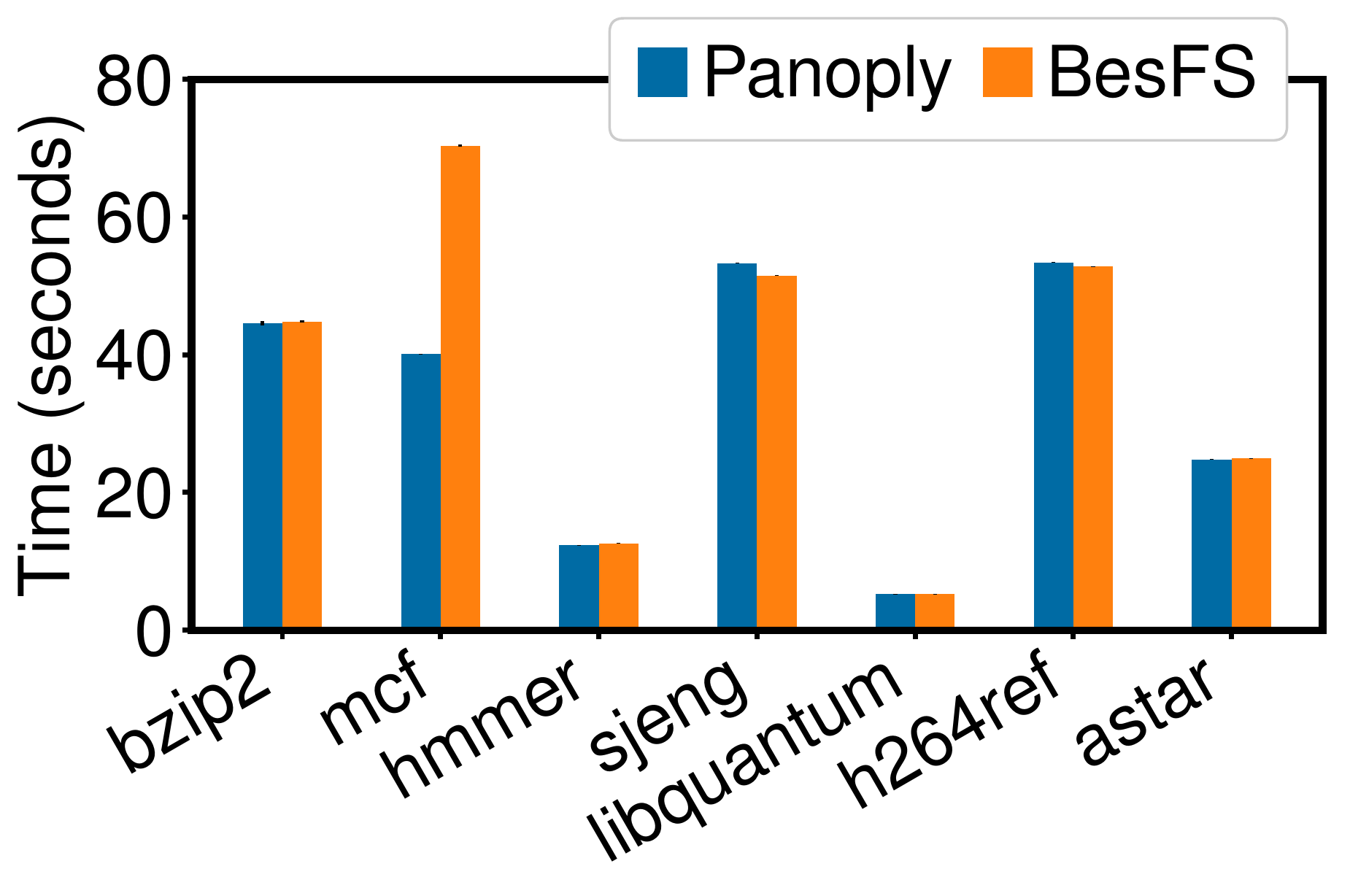, scale=0.22}
\subcaption{SPEC CINT 2006. }	
\label{fig:spec}
\end{minipage}
\begin{minipage}[h]{0.24\textwidth}
\epsfig{file=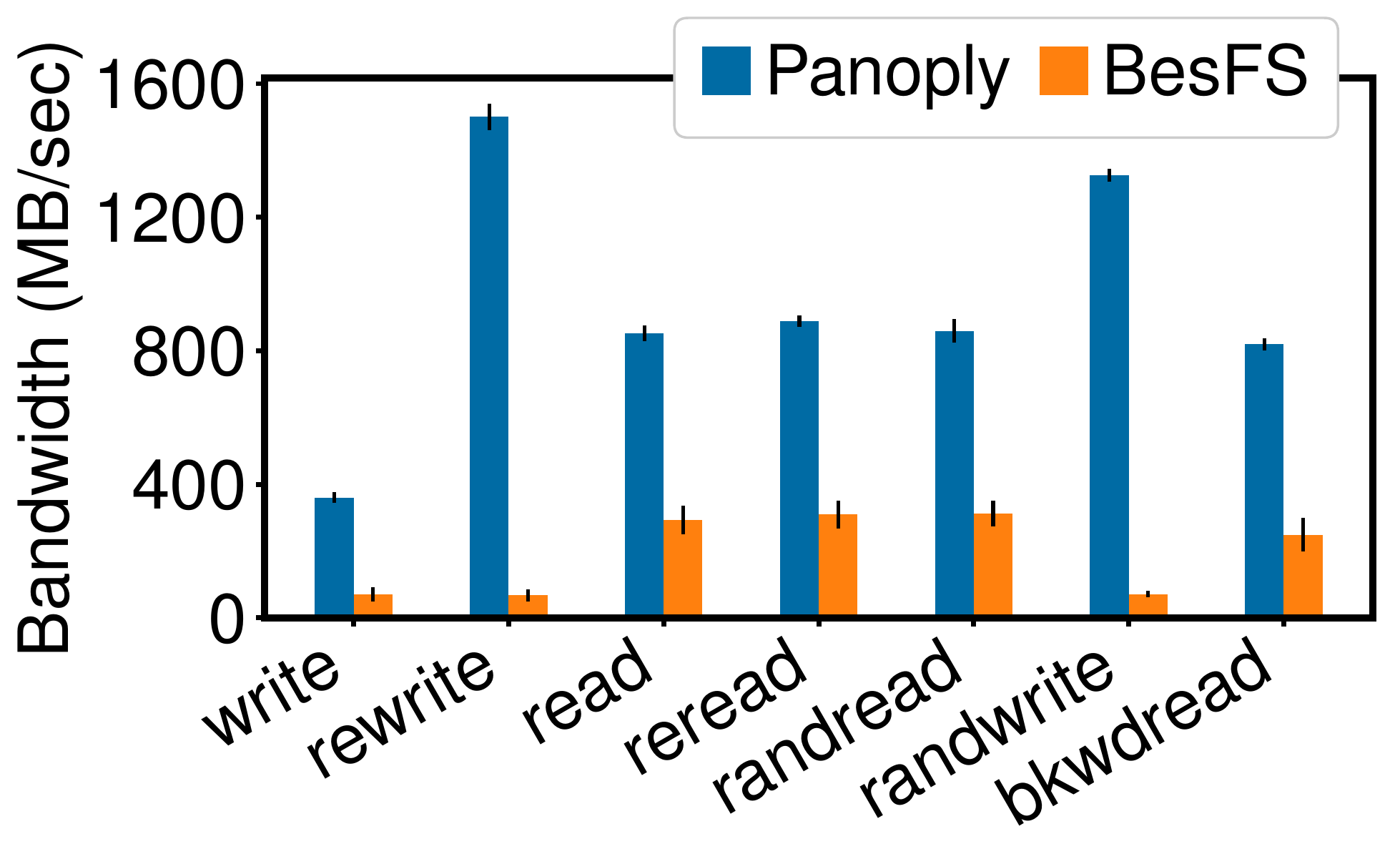, scale=0.22}
\subcaption{IOZone.}	
\label{fig:iozone}
\end{minipage}
\vspace{-5pt}
\caption{\codename Performance on micro-benchmarks, standard CPU, and
IO benchmarks with respect to \panoply. 
(a) Execution overhead for each system call in \fscq.
(b) File operation bandwidth reported by \fscq large IO suite.
(c) Execution overhead on SPEC2006 CPU benchmarks.
(d) File operation bandwidth reported by IOZone benchmarks.}
\label{fig:perf}
\vspace{-15pt}
\end{figure*}

%% file: chapters/perf.tex
\subsection{Performance}
\label{subsec:eval-perf}

\codename is the first formally verified filesystem for SGX. Although
our primary goal is not performance, we report performance on our
benchmarks for completeness. First, we report the per-call latencies
and file access pattern latencies with the \fscq micro-benchmark. Our
main take away from this experiment is that \codename overhead is
dominated by the encryption-decryption of the file content. Next, we
demonstrate this phenomenon systematically by reporting $12.22\%$
overhead and $4.8\times$ bandwidth slowdown on standard CPU (SPEC
CINT2006) and I/O benchmarks respectively. Lastly, we report the
overheads on real-world applications in
Section~\ref{sec:case-studies}. Future optimizations can use \codename
API specification as an oracle for golden implementation.

\paragraph{Micro-benchmarking Single File-related Operations.}
We use \fscq to measure the per-system call overhead of \codename.
Figure~\ref{fig:single} shows that it averages to $3.1\times$. We
observe that read-write operations incur a large overhead. The read
operation is slowed down by $3.7\times$ and create+write is
$5.4\times$ slower because \codename performs page-level AES-GCM
authenticated encryption when the file content is stored on the disk.
Thus, each read and write operation leads to at least a page of
encryption-decryption and integrity computation.

\paragraph{Micro-benchmarking Access Patterns.}
Next, we run all the large tests in \fscq with $8$ KB block size, $1$
KB I/O transfer size, and $1$ MB total file size. \fscq performs a
series of sequential write, sequential read, re-read, random read,
random write, multi-write, and multi-read operations. We perform each
type of operation $100$K times on the files. We observe an average
overhead of $6.7\times$ because of \codename checks.
Figure~\ref{fig:multi} shows the bandwidth for each of these
operations. Sequential access incurs relatively less performance
overhead because they consolidate the page-level encryption-decryption
for every $4$K bytes. Random accesses are more expensive because each
read / write may cause a page-level encryption-decryption. \codename
does not cache page content so re-reads and sequential reads incur
similar overheads.

\paragraph{I/O Intensive Benchmark: IOZone.}
We use IOZone to test \codename for file sizes up to $512$ KB while
varying the record size from $4$ KB to $512$ KB and report the aggregate
performance in Figure~\ref{fig:iozone}. We observe an average of
$4.8\times$ decrease in the IO bandwidth over all the operations.
Write operations are significantly more expensive in comparison to
reads. This is because \codename performs reads over the page for
decrypting the content and then does a write, which requires
encryption. 

\paragraph{CPU Intensive Benchmark: SPEC CINT2006.}
SPEC benchmarks take in a configuration file and optionally an input
file to produce an output file. Figure~\ref{fig:spec} and shows the
performance per-application overhead; the average overhead is
$12.22\%$. $\tt{hmmer}$, $\tt{href}$, $\tt{sjeng}$, and
$\tt{libquantum}$ have relatively less overhead whereas $\tt{astar}$,
$\tt{bzip2}$, and $\tt{mcf}$ exhibit larger overhead. $\tt{astar}$ and
$\tt{mcf}$ use $\tt{fscanf}$ to read the configuration files. Thus,
reading each character read leads to a page read and corresponding
decryption and integrity check. Further, $\tt{astar}$ reads a binary
size of $65$ KB for processing. As shown by our single syscall
measurements (Figure~\ref{fig:single}), reads are expensive. Both
these factors amplify the slowdown for $\tt{astar}$. $\tt{bzip2}$ and
$\tt{mcf}$ output the benchmark results to new files of sizes $274$
and $32$ KB respectively which leads to a slowdown. Specifically,
$\tt{bzip2}$ reads input file in chunks of $5000$ bytes which leads to
a $2$-page read / write and decrypt/encrypt per chunk. Finally,
$\tt{libquantum}$ has the lowest overhead because it does not perform
any file operations. 

%% file: chapters/case-studies.tex
\subsection{Real-world Case Studies}
\label{sec:case-studies}

We showcase the ease of integration and usage of \codename in
real-world enclave programs with two case-studies: 
(a) $4$ applications from \panoply; 
(b) $10$ applications from \privado~\cite{privado} which is built
directly on Intel SGX SDK.

\paragraph{Secure Micron Execution with \panoply.}
We use the $4$ applications from the \panoply paper and evaluate them
under the same workloads~\cite{panoply}. We do not observe any
significant slow-down for OpenSSL($\pm0.2\%$) and Tor nodes
($\pm0.8\%$). Both these applications use file
operations to load configurations (e.g., signing keys, certificates,
node information) only once during their lifetime, while the rest of
the execution does not interact with files. On the other hand, we
observe an overhead of $72.5\%$ for the FreeTDS client.
We attribute this overhead to the nature of the application which
performs file operations for each of the $48$ SQL queries in the
benchmarks. Lastly, we report that the H2O web server logic violates 
\codename safety properties. Specifically, H2O duplicates
the file descriptors across worker threads and concurrently accesses
the file content to be served to the clients. Thus, we deem H2O as
non-compatible with \codename.

\paragraph{Secure Inference with \privado.} 
As a second case study, we integrate \codename with
\privado~\cite{privado}---an SGX-compatible machine-learning framework.
It uses Torch library to infer labels of images from standard datasets
using $10$ well-known deep neural net architectures (LeNet, VGG19,
Wideresnet, Resnet110, Resnext29, AlexNet, Squeezenet, Resnet50,
Inceptionv3, and Densenet). These applications vary from $230$ LOC to
$13.4$ KLOC and have enclave memory footprint between $0.6$ MB to
$392$ MB. We use Cifar-10 and ImageNet datasets, as done in \privado,
where each image is $3.1$~KB and $155.6$~KB respectively. For each of
the application, we integrate \codename interface with $20$ LOC
changes to \privado. We observe an overhead of $\pm1\%$ relative to
the baseline for all the networks and their corresponding datasets. We
see such low overheads because, unlike \panoply, \privado decrypts the
file input after reading it. Thus, the baseline includes the cost of
decryption. In this case, \codename only adds a fixed startup cost of
checks proportional to the number of file operations on the input file
and the number of images in a batch, while keeping the decryption time
constant across both the systems. This shows that \codename is 
compatible and easy to integrate with enclaves which
already use file encryption-decryption.

%% file: chapters/related.tex
\section{Related Work}
\label{sec:related}

We survey the existing SGX defenses including verification as well as
filesystem hardening work in the non-SGX setting.

\paragraph{SGX Attacks \& Defenses.}
\codename ensures the filesystem integrity based on hardware integrity
guarantees of SGX. It assumes the confidentiality of SGX only in one
lemma, i.e., the secrecy of a cryptographic key. This is an important
design choice in light of the side-channels~\cite{cca-sgx, pigeonhole,
branch-shadowing, spectre}. \codename assumes secure hardware
implementation and is agnostic to confidentiality
defenses~\cite{sgx-cache-transmem}.

\paragraph{Filesystem Support in SGX.} 
Ideally, the enclave should not make any assumptions about the
faithful execution on the untrusted calls and should do its due
diligence before using any (implicit or explicit) results of each
untrusted call. The effects of malicious behavior of the OS on the
enclave's execution depends on what counter-measures the enclave has
in place to detect and / or protect against an unfaithful OS.
Currently, the common ways to facilitate the use of filesystem APIs
inside an enclave are (a) port the entire filesystem inside the
enclave~\cite{obliviate,ryoan}; (b) keep the files encrypted outside
the enclave~\cite{panoply, graphene-sgx, scone} and, for each return
parameters, check the data types, bounds on the IO buffers, and valid
value ranges of API specific values (e.g., error codes, flags, and
structures). As one concrete comparison, Intel SGX SDK PFS
Library~\cite{intel-fs} is dedicated solely to the filesystem layer.
Although it leaves the enclave vulnerable to Iago-like attacks as we
showed in Section~\ref{sec:examples}, it is better than approaches
which bloat the TCB to support all syscalls. It is not transparent to
existing legacy applications; the enclave has to use APIs with the
non-standard interface for explicit key management (e.g.,
$\tt{sgx\_fopen\_auto\_key}$) as well as traditional file operations
(e.g., $\tt{sgx\_fopen(filename, mode, key)}$). More importantly,
while these systems reduce the attack surface of file syscall return
value tampering, none of them provably thwart all the attacks in
Section~\ref{sec:attacks}. Other filesystems with untrusted OS in a
non-enclave setting are not formally verified~\cite{sego}.

\paragraph{Verified Guarantees for Enclaves.}
Formal guarantees have been studied for enclaved applications to some
extent. They provide provable confidentiality guarantees for pieces
of code executing inside the enclave. Most notably, Moat~\cite{moat},
/Confidential~\cite{slashconfidential}, and
IMPe~\cite{sgx-policies-oopsla16} formally model various adversary
models in SGX and ensures that the enclave code does not leak
confidential information. These confidentiality efforts are orthogonal
to \codename's integrity goals. Another line of verification research
has focused on certifying the properties of the SGX hardware primitive
itself, which \codename assumes to be correctly implemented. 
Komodo~\cite{komodo-sosp17} is a formally specified and verified
monitor for isolated execution which ensures the confidentiality and
integrity of enclaves. TAP~\cite{tap-ccs17} does formal modeling and
verification to show that SGX and Sanctum~\cite{sanctum} provide
secure remote execution which includes integrity, confidentiality, and
secure measurement. The existing works on verified filesystems do not
reason about an untrusted OS so they cannot be simply added on top of
these enclave systems. \codename is above these hardware
abstractions.

\paragraph{Filesystem Verification.}
Formal verification for large-scale systems such as operating
systems~\cite{sel4-sosp, certikos}, hypervisors\cite{hypervisor-veri},
driver sub-systems~\cite{device-veri} and
user-applications~\cite{ironclad} has been a long-standing area of
research. None of these works consider a Byzantine OS, which leads to
completely different modeling of properties. Filesystem verification
for benign OS, however, is in itself a challenging
task~\cite{fs-veri-too}. This includes building abstract
specifications~\cite{posix-fs-local}, systematically finding
bugs~\cite{fs-mc-errors}, POSIX non-compliance~\cite{sibylfs} in
filesystem implementations, end-to-end verified
implementations~\cite{cogent}, crash consistency~\cite{ferrite}, and
crash recovery~\cite{fscq}.

%% file: chapters/discussion.tex
\section{Discussion}
\label{sec:discussion}

While \besfs has a machine-checked \coq implementation of our 
filesystem API specification, it would be desirable to have
machine-checked enclave-executable code. We believe this is feasible,
in principle, but requires significant advances in state-of-the-art
certified language techniques to become immediately practical. There
are at least three different promising future work directions to
enable certified executable \besfs code: (1) directly certifying the
enclave machine code~\cite{framac}; (2) using a certified compiler to
convert \coq code to machine code~\cite{certicoq}; and (3) using a
simulation proof of $\tt{C}$ or machine code implementation with the
\coq code in the spirit of K~\cite{k-framework}.

\paragraph{Compiling \coq to $\tt{C}$.}
The most promising direction is to have certified compilation from
\coq to $\tt{C}$ code and then from $\tt{C}$ to machine code.
CertiCoq~\cite{certicoq} is a certified compiler from \gallina (\coq)
to CompCert-$\tt{C}$. CompCert~\cite{compcert} is one of the most
mature certified $\tt{C}$ compiler which ensures that the generated
machine code for various processors behaves exactly as prescribed by
the semantics of the source program. With help from the CertiCoq team,
we report that we have successfully compiled \besfs to executable
$\tt{C}$ code. However, we point out that CertiCoq is a very early
stage compiler at present. The produced code is incomplete which
causes segmentation faults. Further, it cannot be interfaced with
external function calls (e.g. system calls) due to missing foreign
function interfaces (FFI). Nonetheless, we expect that as CertiCoq
matures, certified machine code for \besfs (and similar systems) will
become a practical possibility.

\paragraph{Verified Machine Code.}
The second possibility is to verify the machine code directly. Given
that \besfs is written at a higher level of abstraction (\gallina),
our subsequent verification has to reason about the language
abstraction gap between \gallina and machine code. \coq supports
extraction to OCaml, Haskell, Scala, and $\tt{C}$. The most mature
extraction techniques are to OCaml and Haskell, so we tried to port
their runtimes to SGX. For reasons reported in
Section~\ref{sec:extraction}, porting such language runtimes to SGX
certifiably merits a separate research effort in its own right.

\paragraph{Bisimulation.}
A third possibility is a bisimulation of the $\tt{C}$ or machine code
and the \coq code. For maintaining such proofs, when the \besfs
specification expands in the future, the best way is to specify the
operational semantics of the machine code (or $\tt{C}$) and \coq in a
common framework. We believe this is possible but entails significant
future work.

%% file: chapters/conclusion.tex
\section{Conclusion}
\label{sec:conclusion}
\codename is the first formally proved enclave specification and
implementation for integrity-protecting POSIX filesystem API.
\codename API is expressive to support real applications, minimizes
the TCB, and eliminates bugs.

%% file: chapters/ack.tex
\section*{Acknowledgments}
We thank our shepherd Vasileios Kemerlis and the anonymous reviewers
for their feedback; 
Andrew Appel and Olivier Belanger for discussions and help with 
CertiCoq; 
Privado team at Microsoft Research for sharing the torch code for our
case-study;
Shruti Tople, Shiqi Shen, Teodora Baluta, and Zheng Leong Chua for
their help on improving earlier drafts of the paper.
This research was partially supported by a grant from the National
Research Foundation, Prime Ministers Office, Singapore under its
National Cybersecurity R\&D Program (TSUNAMi project, No.
NRF2014NCR-NCR001-21) and administered by the National Cybersecurity
R\&D Directorate. 
This work was funded in part by Yale-NUS College grant
R-607-265-322-121.
This material is in part based upon work supported by the National
Science Foundation under Grant No. DARPA N66001-15-C-4066 and Center
for Long-Term Cybersecurity. Any opinions, findings, and conclusions
or recommendations expressed in this material are those of the
authors and do not necessarily reflect the views of the National
Science Foundation.

%% file: chapters/availability.tex
\section*{Availability}

\codename specification and implementation in \coq is
available at \url{https://shwetasshinde24.github.io/BesFS/}